\documentclass[12pt]{article}
\pdfoutput=1
\usepackage{ae}
\usepackage[T1]{fontenc}
\usepackage[ansinew]{inputenc}
\usepackage{mathrsfs}
\usepackage{amsmath}
\usepackage{amssymb}
\usepackage{graphicx}
\usepackage{color}
\definecolor{darkblue}{cmyk}{0.9,0.9,0,0}
\definecolor{carageen}{RGB}{0,0.55,0}
\usepackage[colorlinks=true,linkcolor=darkblue,citecolor=darkblue,urlcolor=darkblue]{hyperref}
\usepackage{epsfig}
\usepackage{wasysym}

\usepackage{amsmath,amsfonts,amssymb,amsbsy}
\usepackage{graphicx,color}
\usepackage{graphics}
\usepackage{epsfig}
\usepackage{verbatim}
\usepackage{stmaryrd}
\usepackage{epsf}
\usepackage{amsmath, amssymb,graphicx}
\usepackage{epsfig}
\usepackage{verbatim}
\usepackage{amsfonts}
\usepackage{ulem}
\usepackage{cite}

\def\bc{\begin{center}}
\def\ec{\end{center}}

\newcommand{\be}{\begin{equation}}
\newcommand{\ee}{\end{equation}}
\newcommand{\ba}{\begin{eqnarray}}
\newcommand{\ea}{\end{eqnarray}}
\newcommand{\nn}{{\nonumber}}
\newcommand{\beaa}{\begin{eqnarray}}
\newcommand{\eeaa}{\end{eqnarray}}

\newcommand{\dn}{{\,\mathrm{dn}}}
\newcommand{\sn}{{\,\mathrm{sn}}}
\newcommand{\cn}{{\,\mathrm{cn}}}

\definecolor{carageen}{RGB}{0.1,0.7,0.1}

\DeclareFontFamily{OT1}{pzc}{}
\DeclareFontShape{OT1}{pzc}{m}{it}{<-> s * [1.10] pzcmi7t}{}
\DeclareMathAlphabet{\mathpzc}{OT1}{pzc}{m}{it}

\def\({\left(}
\def\){\right)}
\def\[{\left[}
\def\]{\right]}

\def\<{\langle}
\def\>{\rangle}

\def\nref#1{(\ref{#1})}

\begin{document}

\thispagestyle{empty}

\renewcommand{\thefootnote}{\fnsymbol{footnote}}
\setcounter{footnote}{0}
\setcounter{figure}{0}
\begin{center}
$$$$
{\Large\textbf{Semiclassical partition function for strings dual to Wilson loops with small cusps in ABJM}\par}

\vspace{1.0cm}

\textrm{Jerem\'\i as Aguilera-Damia, Diego H. Correa and  Guillermo
A. Silva}
\\ \vspace{1.2cm}
\footnotesize{Instituto de F\'{\i}sica La Plata, CONICET
\\ Departamento de F\'{\i}sica, Universidad Nacional de La Plata
\\ C.C. 67, 1900 La Plata, Argentina}

\par\vspace{1.5cm}

%%%%%%%%%%%%%%%%%%%%%%%%%%%%%%%%%%%%%%%%%%%%%%%%%%%%%%%%%%%%%%%%%%%%%%%%%%%%%%%%%%%%%%%%%%%%%%%%%%%%%%%%%%%%%%
%%%%%%%%%%%%%%%%%%%%%%%%%%%%%%%%%%%%%%%%%%%%%%%%%%%%%%%%%%%%%%%%%%%%%%%%%%%%%%%%%%%%%%%%%%%%%%%%%%%%%%%%%%%%%%
%%%%%%%%%%%%%%%%%%%%%%%%%%%%%%%%%%%%%%%%%%%%%%%%%%%%%%%%%%%%%%%%%%%%%%%%%%%%%%%%%%%%%%%%%%%%%%%%%%%%%%%%%%%%%%

\textbf{Abstract}\vspace{2mm}
\end{center}

\noindent
We compute the 1-loop partition function for strings in $AdS_4\times\mathbb{CP}^3$,
whose worldsheets end along a line with small cusp angles in the boundary of AdS.
We obtain these 1-loop results in terms of the vacuum energy for on-shell modes.
Our results verify the proposal by Lewkowycz and Maldacena in arXiv:1312.5682
for the exact Bremsstrahlung function up to the next to leading order in the strong
coupling expansion. The agreement is observed for cusps distorting either the
1/2 BPS or the 1/6 BPS Wilson line.

%%%%%%%%%%%%%%%%%%%%%%%%%%%%%%%%%%%%%%%%%%%%%%%%%%%%%%%%%%%%%%%%%%%%%%%%%%%%%%%%%%%%%%%%%%%%%%%%%%%%%%%%%%%%%%
%%%%%%%%%%%%%%%%%%%%%%%%%%%%%%%%%%%%%%%%%%%%%%%%%%%%%%%%%%%%%%%%%%%%%%%%%%%%%%%%%%%%%%%%%%%%%%%%%%%%%%%%%%%%%%
%%%%%%%%%%%%%%%%%%%%%%%%%%%%%%%%%%%%%%%%%%%%%%%%%%%%%%%%%%%%%%%%%%%%%%%%%%%%%%%%%%%%%%%%%%%%%%%%%%%%%%%%%%%%%%

\vspace*{\fill}

\setcounter{page}{1}
\renewcommand{\thefootnote}{\arabic{footnote}}
\setcounter{footnote}{0}

\newpage
\tableofcontents

%%%%%%%%%%%%%%%%%%%%%%%%%%%%%%%%%%%%%%%%%%%%%%%%%%%%%%%%%%%%%%%%%%%%%%%%%%%%%%%%%%%%%%%%%%%%%%%%%%%%%%%%%%%%%%
%%%%%%%%%%%%%%%%%%%%%%%%%%%%%%%%%%%%%%%%%%%%%%%%%%%%%%%%%%%%%%%%%%%%%%%%%%%%%%%%%%%%%%%%%%%%%%%%%%%%%%%%%%%%%%
%%%%%%%%%%%%%%%%%%%%%%%%%%%%%%%%%%%%%%%%%%%%%%%%%%%%%%%%%%%%%%%%%%%%%%%%%%%%%%%%%%%%%%%%%%%%%%%%%%%%%%%%%%%%%%

\section{Introduction}

%%%%%%%%%%%%%%%%%%%%%%%%%%%%%%%%%%%%%%%%%%%%%%%%%%%%%%%%%%%%%%%%%%%%%%%%%%%%%%%%%%%%%%%%%%%%%%%%%%%%%%%%%%%%%%
%%%%%%%%%%%%%%%%%%%%%%%%%%%%%%%%%%%%%%%%%%%%%%%%%%%%%%%%%%%%%%%%%%%%%%%%%%%%%%%%%%%%%%%%%%%%%%%%%%%%%%%%%%%%%%
%%%%%%%%%%%%%%%%%%%%%%%%%%%%%%%%%%%%%%%%%%%%%%%%%%%%%%%%%%%%%%%%%%%%%%%%%%%%%%%%%%%%%%%%%%%%%%%%%%%%%%%%%%%%%%

The existence of exact results for vacuum expectation values of  Wilson loops
in supersymmetric gauge theories paved the way for many non-trivial explicit verifications
of the AdS/CFT correspondence. In various cases the strong coupling expansion of exact results was successfully contrasted with explicit 1-loop results for the dual string theory configurations.
For instance, the expectation value of a Wilson line with a small cusp angle along it, can be exactly computed in ${\cal N} = 4$ super Yang-Mills by relating it to the circular Wilson loop \cite{Correa:2012at,Fiol:2012sg} and in ${\cal N}=6$ super Chern-Simons-matter gauge theories by showing its relation to a multiply wound circular Wilson loop \cite{Lewkowycz:2013laa}. The exact results in the former were shown to agree with the 1-loop partition function of the dual string configuration \cite{Drukker:2011za}. However, the exact results for cusped Wilson lines in the latter are yet to be matched with explicit computations in the strong coupling limit.

Three dimensional ${\cal N}=6$ super Chern-Simons-matter is a theory with  $U(N)\times U(N)$ gauge group and
Chern-Simons levels $(k,-k)$. This theory gave rise to another precise AdS/CFT correspondence example, since
it was found to be dual to eleven-dimensional M-theory on $AdS_4\times S^7/Z_k$ \cite{abjm}. Moreover, in the $k\to\infty$ limit, the M-theory reduces to type IIA string theory on $AdS_4\times\mathbb{CP}^3$.
It is therefore interesting to study Wilson loop observables either from the CFT or the AdS point of view,
to understand different regimes and to produce verifications of the AdS/CFT correspondence in cases for
which exact results are available. In particular, the observables we are interested in this work are cusped Wilson lines.

Cusped Wilson lines in ${\cal N}=6$ super Chern-Simons-matter theory have been studied in several works both
from the gauge and the string theory sides \cite{Griguolo:2012iq,Cardinali:2012ru,forini,forini2loop}, and
related to circular Wilson loops in \cite{Lewkowycz:2013laa,Drukker:2009hy,Bianchi:2014laa,Correa:2014aga}.
Wilson lines with a cusp angle $\phi$ in $\mathbb{R}^3$ can be mapped through a conformal transformation to a pair of antiparallel lines in $ S^2\times\mathbb{R}$ separated by an angle $\pi-\phi$ on the sphere. The expectation value of such operator gives rise to the so called cusp anomalous dimension $\Gamma_{\rm cusp}$ defined as
\be
\langle W_{\rm cusp} \rangle = e^{-\Gamma_{\rm cusp}T}\,,
\ee
where $T$ is the temporal extension of the anti-parallel lines along the cylinder. It is clear from this expression that the cusp anomalous dimension is nothing but the potential between the two probes on the sphere. We will be considering Wilson lines that become 1/2 BPS when the cusp angle vanishes. These cusped Wilson loop operators admit a second deformation by an internal space cusp angle $\theta$ \cite{Griguolo:2012iq}, which accounts for a sudden change in coupling with the matter fields.
It has been proven that generalized cusped Wilson lines with $\theta = \pm \phi$ remain BPS \cite{Griguolo:2012iq,Cardinali:2012ru}. Therefore, for $\theta,\phi\ll 1$, the cusp anomalous dimension takes the form
\be
\Gamma_{\rm cusp} \approx (\theta^2-\phi^2) B(\lambda)\,.
\label{BPS}
\ee
The coefficient $B(\lambda)$ is called the Bremsstrahlung function because it appears in the expression
for the energy radiated by an accelerated fundamental charge \cite{Correa:2012at}.

The generalized $\Gamma_{\rm cusp}$ for ABJM theory was computed perturbatively in \cite{Griguolo:2012iq} to 2-loop order. In the strong coupling limit, expectation values of Wilson loops defined for a
curve $\cal C$ can be computed, using the AdS/CFT correspondence, as the partition function for a string worldsheet ending on the curve $\cal C$ located at the boundary. The dual gravity background for ${\cal N}=6$ super Chern-Simons-matter theory  is $AdS_4\times \mathbb{CP}^3$  and the partition function for a cusped line, to subleading order in the strong coupling limit, have been studied in \cite{forini}. The expressions for the Bremsstrahlung function $B(\lambda)$ found in \cite{forini} do not coincide in the geometrical and internal cusps cases. As noted in \cite{forini}, this is in contradiction to the fact that the Wilson loop remains BPS when $\theta = \pm \phi$ \cite{Griguolo:2012iq}. This is  our first motivation for revisiting the 1-loop computation for the partition function of strings in $AdS_4\times \mathbb{CP}^3$ ending in cusped lines.

In \cite{Lewkowycz:2013laa}, a proposal for the exact Bremsstrah\-lung function in ABJM theories in terms of derivatives of a circular Wilson loop expectation value was given
\be
B(\lambda,N) = \frac{1}{4\pi^2} \left.\partial_n \log|\langle W_n \rangle|\right|_{n=1}\,.
\label{mastereq}
\ee
Here $W_n$ is a BPS Wilson loop that winds $n$ times around a maximal circle in the 3-sphere
where the ABJM is defined.  For the case of a geometrical cusp
angle in a 1/6 BPS Wilson line, the 1/6 BPS circular Wilson loop computed in \cite{Kapustin:2009kz,Drukker:2010nc,Klemm:2012ii}
was used in \cite{Lewkowycz:2013laa} to obtain explicit weak and strong coupling expansions
for $B(\lambda)$. For the case of a geometrical cusp angle in a 1/2 Wilson line, the use of (\ref{mastereq}) to compute $B(\lambda,N)$ was slightly discussed in \cite{Lewkowycz:2013laa}. Later on, in \cite{Bianchi:2014laa} explicit weak and strong coupling expansions were written down and a disagreement with the 1-loop strong coupling computation of \cite{forini} was observed. Overcoming such discrepancy has also motivated our work.

The paper is organized as follows: In section \ref{anomaly} we revisit the arguments which allow
to compute the partition function in terms of the vacuum energy density and verify, as expected, that the vacuum
energy density vanishes for the case of a 1/2 BPS straight line in ABJM theory. We then proceed to compute the
correction to the vacuum energy due to small geometrical and internal
cusp angles, using standard perturbation theory. In section \ref{exact} we verify that the proposal
\eqref{mastereq} is correct to  1-loop order in the strong coupling expansion. In section \ref{discussion} we discuss our main results and explain the reasons for the discrepancy with the results in \cite{forini}.

%%%%%%%%%%%%%%%%%%%%%%%%%%%%%%%%%%%%%%%%%%%%%%%%%%%%%%%%%%%%%%%%%%%%%%%%%%%%%%%%%%%%%%%%%%%%%%%%%%%%%%%%%%%%%%
%%%%%%%%%%%%%%%%%%%%%%%%%%%%%%%%%%%%%%%%%%%%%%%%%%%%%%%%%%%%%%%%%%%%%%%%%%%%%%%%%%%%%%%%%%%%%%%%%%%%%%%%%%%%%%
%%%%%%%%%%%%%%%%%%%%%%%%%%%%%%%%%%%%%%%%%%%%%%%%%%%%%%%%%%%%%%%%%%%%%%%%%%%%%%%%%%%%%%%%%%%%%%%%%%%%%%%%%%%%%%

\section{String partition function via vacuum energy}
\label{anomaly}

%%%%%%%%%%%%%%%%%%%%%%%%%%%%%%%%%%%%%%%%%%%%%%%%%%%%%%%%%%%%%%%%%%%%%%%%%%%%%%%%%%%%%%%%%%%%%%%%%%%%%%%%%%%%%%
%%%%%%%%%%%%%%%%%%%%%%%%%%%%%%%%%%%%%%%%%%%%%%%%%%%%%%%%%%%%%%%%%%%%%%%%%%%%%%%%%%%%%%%%%%%%%%%%%%%%%%%%%%%%%%
%%%%%%%%%%%%%%%%%%%%%%%%%%%%%%%%%%%%%%%%%%%%%%%%%%%%%%%%%%%%%%%%%%%%%%%%%%%%%%%%%%%%%%%%%%%%%%%%%%%%%%%%%%%%%%

\subsection{Rescaling of 1-loop operators}

The computation of 1-loop partition functions on a static spacetime with metric $g$ involves
calculating the determinant of some  operator $\mathcal{O}_g$. This determinant is typically computed
by looking for the eigenfuctions of $\mathcal{O}_g$ provided appropriate boundary conditions are given.
This we call the off-shell method. Alternatively, the staticity of spacetime provides an alternative
route: $\mathcal{O}_g$ provides a wave equation whose solutions give rise to a spectrum of states.
This we call  the on-shell method. It is a well known result that the vacuum energy of the on-shell method
coincides with the determinant computed from the off-shell method only when the static spacetime has $g^{00}=1$ \cite{Camporesi:1992wn}. Therefore, in a generic static case the vacuum energy is related to the determinant of $\tilde{\mathcal{O}}_g = \mathcal{M}^{-1}\mathcal{O}_g$ with $\mathcal{M}=g^{00}$.
It was shown in \cite{Schwarz:1979ae} that the determinant of $ \tilde{\mathcal{O}}_g$ differs from the original one by a conformal anomaly type equation (see Appendix \ref{rescaling}). An important observation in \cite{DGT} was to note that the vacuum energy computation can still be applicable to the 1-loop partition function computation if there  is a cancellation among the different anomalies
coming from all the 1-loop operators involved.

As explained in \cite{DGT}, for a string ending on a straight line at the boundary of
$AdS_5 \times S^5$, the 1-loop correction to the partition functions can be obtained
from the vacuum energy because the total anomaly coming from the rescaling of the mode
operators vanishes. In Appendix \ref{rescaling}, we show that this procedure applies for
a string ending on a straight line at the boundary of $AdS_4 \times \mathbb{CP}^3$.
Moreover, the cancellation of anomalies also occurs for a general cusped string configuration
with both geometrical and internal cusps. Therefore, we will use the vacuum energy method to
obtain the corresponding determinants for the 1-loop correction to the partition function.

%%%%%%%%%%%%%%%%%%%%%%%%%%%%%%%%%%%%%%%%%%%%%%%%%%%%%%%%%%%%%%%%%%%%%%%%%%%%%%%%%%%%%%%%%%%%%%%%%%%%%%%%%%%%%%
%%%%%%%%%%%%%%%%%%%%%%%%%%%%%%%%%%%%%%%%%%%%%%%%%%%%%%%%%%%%%%%%%%%%%%%%%%%%%%%%%%%%%%%%%%%%%%%%%%%%%%%%%%%%%%
%%%%%%%%%%%%%%%%%%%%%%%%%%%%%%%%%%%%%%%%%%%%%%%%%%%%%%%%%%%%%%%%%%%%%%%%%%%%%%%%%%%%%%%%%%%%%%%%%%%%%%%%%%%%%%

\subsection{String ending on a straight line}

%%%%%%%%%%%%%%%%%%%%%%%%%%%%%%%%%%%%%%%%%%%%%%%%%%%%%%%%%%%%%%%%%%%%%%%%%%%%%%%%%%%%%%%%%%%%%%%%%%%%%%%%%%%%%%
%%%%%%%%%%%%%%%%%%%%%%%%%%%%%%%%%%%%%%%%%%%%%%%%%%%%%%%%%%%%%%%%%%%%%%%%%%%%%%%%%%%%%%%%%%%%%%%%%%%%%%%%%%%%%%
%%%%%%%%%%%%%%%%%%%%%%%%%%%%%%%%%%%%%%%%%%%%%%%%%%%%%%%%%%%%%%%%%%%%%%%%%%%%%%%%%%%%%%%%%%%%%%%%%%%%%%%%%%%%%%

Expanding the Lagrangian for a straight Wilson line without cusps to second order gives the following
1-loop contribution to the partition function \cite{forini}
\be
Z_{1-loop}=\frac{\prod_j\det^{1/2}(i\tilde\gamma^a{D}_a-\tilde\gamma_* m_j)}
{\det^{6/2}(-\nabla^2)\det^{1/2}(-\nabla^2+R^{(2)}+4)\det^{1/2}(-\nabla^2+2)}\,.
\label{straightz}
\ee
here $\tilde\gamma^a$ represent 2-dimensional curved gamma matrices in the worldsheet, $\tilde \gamma_*=\frac12\varepsilon^{ab}\gamma_{ab}$ is the curved covariant chiral matrix and $D_a$ the corresponding spinorial covariant derivative (see Appendix \ref{classical} for details).
In the absence of cusps, the induced metric on the string worldsheet is an $AdS_2$ geometry, with scalar curvature $R^{(2)}=-2$. The fermionic masses $m_j$ ($j=1,\ldots,8$) in the worldsheet
Dirac operators are related to the eigenvalues of the matrix $M_F$ arising from the Green-Schwarz Lagrangian, which is given in terms of  flat ten dimensional Dirac matrices $\Gamma$ by
\be
M_F=\frac{i}{4}\left[
\left(\Gamma_{49}-\Gamma_{57}+\Gamma_{68}\right)\Gamma_{11}-3\Gamma_{0123}\right]\,.
\ee
Labeling the 32 components spinors as $\Psi_{s_1,s_2,s_3,s_4}$, with $\{s_1,s_2,s_3,s_4\}$ the
eigenvalues of the commuting set of gamma matrices $\{i\Gamma_{49},i\Gamma_{57},i\Gamma_{68},i\Gamma_{23}\}$, using the kappa symmetry gauge fixing $\frac{1}{2}\left(1+\Gamma_{01}\Gamma_{11}\right)\Psi=\Psi$ and an appropriate representation for the Dirac matrices, allow to decompose the 32-component Majorana spinor into eight 2-component Majorana spinors with {\it chiral} mass terms $m_j$. The $m_j$ appearing in \eqref{straightz} are given by
\be
m_j=3s_1s_2s_3+s_1-s_2+s_3\,, \qquad  \quad (s_i=\pm1)\,,
\ee
The result is: two fermions being massless, three with mass $m_F=1$ and three with $m_F=-1$. Naively,
one would have expected a fermionic spectrum with two massless modes, four $m_F=1$ modes and two $m_F=-1$ modes (or some other combination) to match the bosonic spectrum made of six massless and two $m_B^2=2$ bosonic modes, in order to satisfy the $AdS$ supersymmetric relation $m^2_B=m_F^2-m_F$ \cite{Sakai:1984vm}. However, as pointed out in \cite{DGT}, the relation  $m^2_B=m_F^2-m_F$ is valid for ${\cal N} =1$ supersymmetry multiplets, in the case of extended supersymmetry in $AdS_2$, the relation $m^2_B=m_F^2+m_F$ is also possible.

We will compute the on-shell vacuum energy  instead of (\ref{straightz}), therefore we will be computing
the ratio of determinants but with bosonic operators scaled by $g^{1/2}$ and  fermionic
operators are scaled by $g^{1/4}$. Nonetheless, as explained in the previous subsection, these two quantities coincide (see Appendix \ref{rescaling}).

To compute the vacuum energy we have to sum over the on-shell oscillator frequencies. For scalar and fermion fields of masses $m_B$ and $m_F$, the  frequencies were computed in \cite{Sakai:1984vm}
\begin{alignat}{2}
\omega_n^B &= n+\tfrac{1}{2}\left(1+\sqrt{1+4m_B^2}\right)\,.
\label{k0freqbos}
\\
\omega_n^F &=
\left\{\begin{array}{c}
n-m_F+\frac{1}{2}\,, \quad \textrm{ if $m_F<\frac{1}{2}$}\,, \\
n+m_F+\frac{1}{2}\,, \quad \textrm{ if $m_F>\frac{1}{2}$}\,.
\end{array} \right.
\label{k0freqfer}
\end{alignat}
where $n=0,1,2\ldots$. The sums of such frequencies are divergent but can be regularized via analytical continuation using the Hurwitz zeta function
\be
\zeta_H(s,x) = \sum_{n=0}^{\infty}(n+x)^{-s}\,.
\ee
In the present case, we only need the Hurwitz zeta function evaluated at $s=-1$, the result is regular and given by
\be
\zeta_H(-1,x)=-\frac12\left(x^2-x+\frac16\right)\,.
\ee
So, for the case of a straight Wilson line we obtain
\be
\frac{1}{T}\log Z_{1-loop} = \frac{1}{2}\left(6 \zeta_H(-1,\tfrac{3}{2}) + 2 \zeta_H(-1,\tfrac{1}{2}) - 2 \zeta_H(-1,2) - 6 \zeta_H(-1,1)\right) = 0\,.
\label{vanishing}
\ee
The vanishing of the 1-loop contribution to the partition function was the expected result,
because the string ending on a straight line is 1/2 BPS and preserves pure Poincare supercharges.

In \cite{forini}, a non-vanishing 1-loop partition function was obtained for the string ending
on the straight line. Presumably the non-vanishing of the  result has its origin in the
simplification between the determinants from the massless fermion fields and two of the massless scalar fields (see \eqref{geomZ}). Omitting in \eqref{vanishing} the modes that were canceled in \cite{forini} we would have obtained
\be
\frac{1}{2}\left(6 \zeta_H(-1,\tfrac{3}{2}) - 2 \zeta_H(-1,2) - 4 \zeta_H(-1,1)\right) = -\frac{1}{8}\,,
\ee
which is the result reported in  eq. (B.62) in \cite{forini}.
The computation \eqref{k0freqbos}-\eqref{vanishing} shows, as expected, that the determinants of
fermionic and bosonic massless modes do not cancel each other when
computed with  different boundary conditions. This is indeed the case if the excitations are to fit in a supersymmetric multiplet \cite{Sakai:1984vm}.

%%%%%%%%%%%%%%%%%%%%%%%%%%%%%%%%%%%%%%%%%%%%%%%%%%%%%%%%%%%%%%%%%%%%%%%%%%%%%%%%%%%%%%%%%%%%%%%%%%%%%%%%%%%%%%
%%%%%%%%%%%%%%%%%%%%%%%%%%%%%%%%%%%%%%%%%%%%%%%%%%%%%%%%%%%%%%%%%%%%%%%%%%%%%%%%%%%%%%%%%%%%%%%%%%%%%%%%%%%%%%
%%%%%%%%%%%%%%%%%%%%%%%%%%%%%%%%%%%%%%%%%%%%%%%%%%%%%%%%%%%%%%%%%%%%%%%%%%%%%%%%%%%%%%%%%%%%%%%%%%%%%%%%%%%%%%

\subsection{String ending on a line with a geometrical cusp}
\label{geometrical}

%%%%%%%%%%%%%%%%%%%%%%%%%%%%%%%%%%%%%%%%%%%%%%%%%%%%%%%%%%%%%%%%%%%%%%%%%%%%%%%%%%%%%%%%%%%%%%%%%%%%%%%%%%%%%%
%%%%%%%%%%%%%%%%%%%%%%%%%%%%%%%%%%%%%%%%%%%%%%%%%%%%%%%%%%%%%%%%%%%%%%%%%%%%%%%%%%%%%%%%%%%%%%%%%%%%%%%%%%%%%%
%%%%%%%%%%%%%%%%%%%%%%%%%%%%%%%%%%%%%%%%%%%%%%%%%%%%%%%%%%%%%%%%%%%%%%%%%%%%%%%%%%%%%%%%%%%%%%%%%%%%%%%%%%%%%%

Let us now consider a string worldsheet ending on a cusped line with geometrical cusp angle $\phi$
and vanishing internal  cusp angle $\theta =0$ (see Appendix \ref{classical}). The 1-loop contribution to
partition function is still given by (\ref{straightz}), but now covariant derivatives and scalar curvature
are those of the induced worldsheet  metric
\be
ds^2 = \frac{1-k^2}{{\rm cn}^2 (\sigma|k^2)}\left(-d\tau^2+d\sigma^2 \right)\,,\qquad{\rm with}~~~ -K(k^2) <\sigma < K(k^2)\,.
\label{inducedmetric}
\ee
Here ${{\rm cn}(\sigma|k^2)}$ is a Jacobi elliptic function and $K(k^2)$ is the complete elliptic integral of the first kind \cite{Drukker:2011za}. The parameter $k^2$ relates to the geometrical cusp angle in the classical solution. For small cusp angles the relation is $\phi \approx \pi k$. To simplify the notation we shall omit the dependence on $k^2$ in all the elliptic functions. The scalar curvature for the metric \eqref{inducedmetric} is
\be
R^{(2)}=-2\left(1+\frac{k^2}{1-k^2}\textrm{cn}^4(\sigma)\right)\,.
\label{R2}
\ee

As we have done for the straight line, we now proceed to study the on-shell vacuum energy which computes the ratio of determinants of the rescaled operators. After Fourier transforming the time dependence, we obtain the following operators:

\noindent $\bullet$ The eight fermions satisfy\footnote{We have absorbed the chiral matrix present in \eqref{straightz} by a redefinition of the fermion fields and the 2-dimensional gamma matrices.}
\be
\left(i\left(\partial_{\sigma}+\tfrac{{\rm sn}(\sigma){\rm dn}(\sigma)}{2 {\rm cn}(\sigma)}\right)\gamma^1
+\omega_n\gamma^0-m_F\frac{\sqrt{1-k^2}}{{\rm cn}(\sigma)}\right)\Psi_n = 0\,.
\label{gfermop}
\ee
here $\gamma^a$ are 2d flat gamma matrices. Six of the fermions have $m_F=\pm 1$ and the other two have $m_F=0$.

\noindent $\bullet$ Seven of the scalar modes satisfy
\be
\left(\partial_{\sigma}^2 + \omega_n^2 -\frac{m_B^2(1-k^2)}{{\rm cn}^2(\sigma)}\right)\phi_n=0\,,
\label{gbosop1}
\ee
with six modes having $m_B^2=0$ and one $m_B^2=2$. The eighth scalar field equation, whose potential
depends on the scalar curvature $R^{(2)}$, results in
\be
\left(\partial_{\sigma}^2 + \omega_n^2 - \frac{2(1-k^2)}{{\rm cn}^2(\sigma)} + 2 k^2{\rm cn}^2(\sigma)\right)\phi_n =0\,.
\label{gbosopR}
\ee
For $k=0$ these operators reduce to either massless or massive Dirac and Klein-Gordon fields in $AdS_2$.

We now quote from \cite{Sakai:1984vm} the pertaining solutions to $AdS_2$ since we will exploit them in the next section.

\noindent {\sf Fermions}:  depending on the value of the mass, two possible solutions exist comprising supersymmetric multiplets:\footnote{The choice of gamma matrices in \eqref{gfermop} is $\gamma^a=(\sigma^1,i\sigma^3)$.}

\noindent (I) For  $m_F<1/2$, we denote the fermion by $\Psi^{T}_{m,n} = (\psi^1_{m,n},\psi^2_{m,n})$ and one has
\be
\begin{array}{c}
\psi^1_{m,n}(\sigma)= \frac{\left[n!\Gamma(n-2m+1)\right]^{\frac12}}{2^{-m}\Gamma(n-m+\frac12)}\cos^{-m+\frac12}(\sigma)\cos(\frac{\sigma}{2}+\frac{\pi}{4})
P_n^{(-m+\frac12 , -m-\frac12)}(\sin\sigma)\,,\\
\psi^2_{m,n}(\sigma)= -\frac{\left[n!\Gamma(n-2m+1)\right]^{\frac12}}{2^{-m}\Gamma(n-m+\frac12)}\cos^{-m+\frac12}(\sigma)\sin(\frac{\sigma}{2}+\frac{\pi}{4})
P_n^{(-m-\frac12 , -m+\frac12)}(\sin\sigma)\,,
\end{array}
\label{feigen}
\ee
here $P_n$ denotes the Jacobi polynomials.

\noindent (II) For $m_F>-1/2$, we denote the fermion as ${\rm X}^{T}_{m,n} = (\chi^1_{m,n},\chi^2_{m,n})$ and
the components result
\be
\begin{array}{c}
\chi^1_{m,n}(\sigma)= \frac{\left[n!\Gamma(n+2m+1)\right]^{\frac12}}{2^{m}\Gamma(n+m+\frac12)}\cos^{m+\frac12}(\sigma)\sin(\frac{\sigma}{2}+\frac{\pi}{4})
P_n^{(m-\frac12 , m+\frac12)}(\sin\sigma)\,,\\
\chi^2_{m,n}(\sigma)= \frac{\left[n!\Gamma(n+2m+1)\right]^{\frac12}}{2^{m}\Gamma(n+m+\frac12)}\cos^{m+\frac12}(\sigma)\cos(\frac{\sigma}{2}+\frac{\pi}{4})
P_n^{(m+\frac12 , m-\frac12)}(\sin\sigma)\,.
\end{array}
\label{feigen2}
\ee
The frequency spectra for these wavefunctions is given by (\ref{k0freqfer}) and their normalizations are such that
\be
\int^{\frac{\pi}{2}}_{-\frac{\pi}{2}} \frac{d\sigma}{\cos\sigma}
\Psi_{m,n}^\dagger \Psi_{m,n'} =
\int^{\frac{\pi}{2}}_{-\frac{\pi}{2}} \frac{d\sigma}{\cos\sigma}
{\rm X}_{m,n}^\dagger {\rm X}_{m,n'} =\delta_{n n'}\,.
\ee

~

\noindent {\sf Bosons}:
The solutions to the Klein-Gordon equation in $AdS_2$ are given by
\be
\phi_{\lambda,n}(\sigma)=\frac{\sqrt{n!\Gamma(n+2\lambda)}}{2^{\lambda+1}\Gamma(n+\lambda+\frac12)(n+\lambda)}
 \cos^\lambda\!\sigma\, P_n^{(\lambda-\frac12 , \lambda-\frac12)}(\sin\sigma)\,,
\label{beigen}
\ee
with frequencies $\omega_n=n+\lambda$ and $\lambda=\frac12(1+\sqrt{1+4m^2})$  (cf. \eqref{k0freqbos}). These wavefunctions are
normalized to satisfy
\be
\int^{\frac{\pi}{2}}_{-\frac{\pi}{2}} d\sigma
\phi_{\lambda,n}^* \phi_{\lambda,n'} =\delta_{n n'}\,.
\ee
All these modes satisfy Dirichlet boundary conditions.

For the massless scalar equation ($\lambda=1$) a second solution exist satisfying Neumann boundary
conditions at $\sigma=\pm\pi/2$ \cite{Sakai:1984vm}
\be
 \phi_{0,n}(\sigma)=\frac{\sqrt{n!(n-1)!}}{\Gamma(n+\frac12)2n}
P_n^{(-\frac12 , -\frac12)}(\sin\sigma)\,,
\label{neumann}
\ee
and the corresponding frequencies are $\omega_n=n$.

However, we are considering string configurations that are sitting at a specific point in internal space $\mathbb{CP}^3$ and therefore we have to impose Dirichlet boundary conditions for the corresponding 6 massless scalar modes. Thus, we use the modes given in \eqref{beigen}. For the massless fermionic
excitations the two possibilities $\Psi_{0,n}$ or ${\rm X}_{0,n}$ are possible in principle. However, only for a specific choice the excitations would conform an ${\cal N}=6$ multiplet. We will come back to this issue later on.

For 1/6 BPS straight Wilson line, the dual configuration is a smearing of strings along a $\mathbb{CP}^1\subset\mathbb{CP}^3$ \cite{DPY,Chen:2008bp,Rey:2008bh} and then two of the massless scalar modes should satisfy Neumann boundary conditions. If we were to repeat the computation done
in \eqref{vanishing} with 2 of the massless scalar modes given by \eqref{neumann}, the total 1-loop vacuum energy would continue to vanish because $\sum(n+1)=\sum n$. In section \ref{discussion} we consider the possibility of using modes \eqref{neumann} the study of smeared Wilson loops deformed with
a geometrical cusp.

~

For $k\neq0$ equations (\ref{gfermop})-(\ref{gbosopR}) can be solved in terms eigenfunctions of the Lam\'e equation. However, and since we are only interested in the leading order of small $k$ limit, we shall simply take the $AdS_2$ eigenfunctions (\ref{feigen}),(\ref{feigen2}) and (\ref{beigen}) and use standard perturbation theory.

%%%%%%%%%%%%%%%%%%%%%%%%%%%%%%%%%%%%%%%%%%%%%%%%%%%%%%%%%%%%%%%%%%%%%%%%%%%%%%%%%%%%%%%%%%%%%%%%%%%%%%%%%%%%%%
%%%%%%%%%%%%%%%%%%%%%%%%%%%%%%%%%%%%%%%%%%%%%%%%%%%%%%%%%%%%%%%%%%%%%%%%%%%%%%%%%%%%%%%%%%%%%%%%%%%%%%%%%%%%%%
%%%%%%%%%%%%%%%%%%%%%%%%%%%%%%%%%%%%%%%%%%%%%%%%%%%%%%%%%%%%%%%%%%%%%%%%%%%%%%%%%%%%%%%%%%%%%%%%%%%%%%%%%%%%%%

\subsubsection{Small cusp}

%%%%%%%%%%%%%%%%%%%%%%%%%%%%%%%%%%%%%%%%%%%%%%%%%%%%%%%%%%%%%%%%%%%%%%%%%%%%%%%%%%%%%%%%%%%%%%%%%%%%%%%%%%%%%%
%%%%%%%%%%%%%%%%%%%%%%%%%%%%%%%%%%%%%%%%%%%%%%%%%%%%%%%%%%%%%%%%%%%%%%%%%%%%%%%%%%%%%%%%%%%%%%%%%%%%%%%%%%%%%%
%%%%%%%%%%%%%%%%%%%%%%%%%%%%%%%%%%%%%%%%%%%%%%%%%%%%%%%%%%%%%%%%%%%%%%%%%%%%%%%%%%%%%%%%%%%%%%%%%%%%%%%%%%%%%%

Turning on $k$ not only perturbs the constant mass in the  Klein-Gordon and Dirac equations in $AdS_2$ by
adding a potential, but also modifies the range of the coordinate $\sigma$. In order to avoid
a $k$-dependent  coordinate range,  it is convenient to define
\be
\tilde\sigma =\frac{\pi \sigma}{2{K}}\,,\qquad\tilde\sigma\in (-\tfrac{\pi}{2},\tfrac{\pi}{2})\,.
\label{rescale}
\ee
and introduce accordingly rescaled frequencies $\omega=\frac{\pi \tilde{\omega}}{2{K}}$.
With these definitions, for instance, the equation (\ref{gbosop1}) becomes
\begin{alignat}{2}
\left(\partial_{\tilde\sigma}^2 + \tilde\omega_n^2 -\left(\tfrac{2K}{\pi}\right)^2\frac{m_B^2(1-k^2)}{{\rm cn}^2(\tfrac{2K}{\pi}\tilde\sigma)}\right)\phi_n &=0 \nn
\\
\overset{\mathcal{O}(k^2)}{\longrightarrow}\left(\partial_{\tilde\sigma}^2 + \tilde\omega_n^2 -\frac{m_B^2}{\cos\tilde\sigma}+k^2\frac{m_B^2}{2}+{\cal O}(k^4)\right)\phi_n &=0
\,.
\end{alignat}

The perturbed frequencies are
\be
\tilde\omega_n^2 = (n + \lambda)^2
- k^2  \int_{-\tfrac{\pi}{2}}^{\tfrac{\pi}{2}}d\tilde{\sigma}\phi_{\lambda,n}^*%(\tilde{\sigma})
\frac{m_B^2}{2} \phi_{\lambda,n}%(\tilde{\sigma})
=  (n + \lambda)^2 - k^2 \frac{m_B^2}{2}\,.
\label{scalarfreq0}
\ee

Thus, the frequencies of the 6 massless modes becomes
\be
\omega_n=(n+1)\left(1-\frac{k^2}{4}\right)+{\cal O}(k^4)\,,
\ee
and each of them contributes to the vacuum energy with
\be
\frac{1}{2}\sum_{n=0}^{\infty}\omega_n
=\frac{1}{2}\zeta_H(-1,1)\left(1-\frac{k^2}{4}\right) +{\cal O}(k^4)\,.
\label{k2bm0}
\ee

For the perturbation of one of the $m_B^2=2$ scalar modes
\be
\omega_n=(n+2)\left(1-\frac{k^2}{4}\right)-\frac{k^2}{2}\frac{1}{n+2} +{\cal O}(k^4)\,,
\ee
which contributes to the vacuum energy with
\be
\frac{1}{2}\sum_{n=0}^{\infty}\omega_n = \frac{1}{2}\zeta_H(-1,2)\left(1-\frac{k^2}{4}\right)
-\frac{k^2}{4}\zeta_H(1,2) + {\cal O}(k^4)\,.
\label{k2bm2}
\ee
The term $\zeta_H(1,2)=\zeta_H(1,1)-1$ originated from the second term in (\ref{omegan}) is divergent but, as we will see in a moment,  it cancels against divergencies coming from  other modes.

For the last scalar mode, the equation (\ref{gbosopR}) becomes
\be
\left(\partial_{\tilde{\sigma}}^2 + \tilde{\omega}^2_n -\frac{2}{\cos^2\tilde\sigma} + k^2\left(1+2\cos^2\tilde{\sigma}\right)+{\cal O}(k^4)\right)\phi_n=0\,,
\ee
where
\be
\tilde\omega_n^2 = (n+2)^2 -k^2\int_{-\tfrac{\pi}{2}}^{\tfrac{\pi}{2}}d\tilde{\sigma}\phi_{2,n}^* (1+2\cos^2\tilde{\sigma})\phi_{2,n}
=(n+2)^2 -k^2\frac{2(n+2)^2}{(n+1)(n+3)}\,,
\ee
which implies that
\be
\omega_n=(n+2)\left(1-\frac{k^2}{4}\right) - \frac{k^2}{n+1}+\frac{k^2}{(n+1)(n+3)}+{\cal O}(k^4)\,.
\label{omegan}
\ee
The contribution of this mode to the vacuum energy is then
\be
\frac{1}{2}\sum_{n=0}^\infty \omega_n =
\frac{1}{2}\zeta_H(-1,2)\left(1-\frac{k^2}{4}\right) -\frac{k^2}{2} \zeta_H(1,1) + \frac{3k^2}{8}+{\cal O}(k^4)\,.
\label{bcont1}
\ee

~

Let us now turn to the fermionic fluctuations. Expanding equation (\ref{gfermop}) we get
\be
\left(i\left(\partial_{\tilde\sigma}+\tfrac{1}{2}\tan\tilde\sigma\right)\gamma^1
+\tilde\omega_n\gamma^0-\frac{m_F}{\cos\tilde\sigma}
+i\frac{k^2}{8}\sin2\tilde\sigma\gamma^1+ \frac{k^2}{4}
m_F\cos\tilde\sigma + {\cal O}(k^4)
\right)\Psi_n = 0\,.
\label{gfermopexpanded}
\ee
Now, using the $k=0$ eigenfunctions, the perturbed frequencies are
\begin{alignat}{2}
\tilde\omega_n = n - m_F +\tfrac{1}{2} & - i\frac{k^2}{8}
\int^{\frac{\pi}{2}}_{-\frac{\pi}{2}} \frac{d\tilde\sigma}{\cos\tilde\sigma}
\Psi_{m_F,n}^\dagger \gamma^{01} \Psi_{m_F,n} \sin2\tilde\sigma\nn\\
& - \frac{k^2}{4}\int^{\frac{\pi}{2}}_{-\frac{\pi}{2}} \frac{d\tilde\sigma}{\cos\tilde\sigma}
\Psi_{m_F,n}^\dagger \gamma^{0} \Psi_{m_F,n} m_F\cos\tilde\sigma\,,
\end{alignat}
for the $m_F<1/2$ fermionic modes or
\begin{alignat}{2}
\tilde\omega_n = n + m_F +\tfrac{1}{2} & - i\frac{k^2}{8}
\int^{\frac{\pi}{2}}_{-\frac{\pi}{2}} \frac{d\tilde\sigma}{\cos\tilde\sigma}
{\rm X}_{m_F,n}^\dagger \gamma^{01} {\rm X}_{m_F,n} \sin2\tilde\sigma\nn\\
& - \frac{k^2}{4}\int^{\frac{\pi}{2}}_{-\frac{\pi}{2}} \frac{d\tilde\sigma}{\cos\tilde\sigma}
{\rm X}_{m_F,n}^\dagger \gamma^{0} {\rm X}_{m_F,n} m_F\cos\tilde\sigma\,,
\end{alignat}
for the $m_F>-1/2$ fermionic modes.

The contribution to the perturbation from the spin connection vanishes because
\be
\Psi_{m_F,n}^\dagger \gamma^{01} \Psi_{m_F,n} = {\rm X}_{m_F,n}^\dagger \gamma^{01} {\rm X}_{m_F,n} = 0\,.
\ee

Computing the remaining integral, for $m_F = \pm 1$, we get
\be
\tilde\omega_n = n + \tfrac{3}{2} -\frac{k^2}8 \frac{2n+3}{(n+2)(n+1)}\,,
\,\, \Rightarrow\,\, \omega_n = (n + \tfrac{3}{2})\left(1-\frac{k^2}{4}\right) -\frac{k^2}4 \frac{1}{n+1}+\frac{k^2}8 \frac{1}{(n+2)(n+1)}\,.
\ee
Then, each of these modes contributes to the vacuum energy with
\be
\frac{1}{2}\sum_{n=0}^{\infty} \omega_n = \frac{1}{2}\zeta_H(-1,\tfrac{3}{2})\left(1-\frac{k^2}{4}\right)
- \frac{k^2}{8}\zeta_H(1,1) +\frac{k^2}{16}\,.
\label{fcont1}
\ee

Finally, for the $m_F=0$ fermionic modes $\tilde\omega = n +\tfrac{1}{2}$ and the correction to the frequencies comes entirely from the change in the range of the variable $\sigma$. Their contribution to the vacuum energy is
\be
\frac{1}{2}\sum_{n=0}^{\infty}\omega_n =\frac{1}{2}\zeta_H(-1,\tfrac{1}{2})\left(1-\frac{k^2}{4}\right) +{\cal O}(k^4)\,.
\label{fcont0}
\ee

Now we collect all the contributions to evaluate the 1-loop partition function. There are 2 massless fermionic fluctuations  (\ref{fcont0}), 6 fermionic fluctuations with masses $m_F =\pm 1$  (\ref{fcont1}), 6 massless scalar fluctuations (\ref{k2bm0}) and 2 scalar fluctuations with $m_B^2 =2$ (\ref{k2bm2}) and (\ref{bcont1}),
\begin{align}
\frac{1}{T}\log{Z_{1-loop}}=& 6\left[\frac{1}{2}\zeta_H(-1,\tfrac{3}{2})\left(1-\frac{k^2}{4}\right) -\frac{k^2}{8}\zeta_H(1,1)+\frac{k^2}{16}\right]%\nn\\
+\frac{2}{2} \zeta_H(-1,\tfrac{1}{2})\left(1-\frac{k^2}{4}\right) \nn\\
&-\frac{6}{2}\zeta_H(-1,1)\left(1-\frac{k^2}{4}\right) -\left[\frac{1}{2}\zeta_H(-1,2)\left(1-\frac{k^2}{4}\right)-\frac{k^2}{2}\zeta_H(1,1)+\frac{3}{8}k^2\right]\nn\\
&-\left[\frac12\zeta_H(-1,2)\left(1-\frac{k^2}{4}\right)-\frac{k^2}{4}\zeta_H(1,2)\right]
%\nn\\
=-\frac{k^2}{4}\,.
\end{align}
Recall that, in this limit, $k^2=\frac{\phi^2}{\pi^2}$ and the 1-loop anomalous dimension results
\be
\Gamma_{\rm cusp}^{1-loop}=\frac{\phi^2}{4\pi^2} +{\cal O}(\phi^4)\,.
\ee
This result differs from the one in \cite{forini}.

~

At this point, and to verify that the vacuum energy computation gives the correct partition function contribution, we can use the above results to reproduce the Bremsstrahlung function to 1-loop order
for a string in $AdS_5 \times S^5$. In that case, all the fermionic fluctuations had masses $m_F =\pm 1$,
5 scalar fluctuations were  massless and 3 had $m_B^2 =2$, two contributing with (\ref{k2bm2}) and the other with (\ref{bcont1}). Thus, in that case the 1-loop partition function is
\begin{align}
\frac{1}{T}\log{Z_{1-loop}}=& 8\left[\frac{1}{2}\zeta_H(-1,\tfrac{3}{2})\left(1-\frac{k^2}{4}\right) -\frac{k^2}{8}\zeta_H(1,1)+\frac{k^2}{16}\right]
\nn\\
& -\frac{5}{2}\zeta_H(-1,1)\left(1-\frac{k^2}{4}\right)
-\left[\frac{1}{2}\zeta_H(-1,2)\left(1-\frac{k^2}{4}\right)-\frac{k^2}{2}\zeta_H(1,1)+\frac{3}{8}k^2\right]
\nn\\
&
-2\left[\frac12\zeta_H(-1,2)\left(1-\frac{k^2}{4}\right)-\frac{k^2}{4}\zeta_H(1,2)\right]
=-\frac{3 k^2}{8}\,,
\end{align}
which leads to
\be
\Gamma_{\rm cusp}^{1-loop}=\frac{3\phi^2}{8\pi^2}\,, \qquad{\rm for}\qquad AdS_5 \times S^5\,,
\ee
in perfect agreement with the 1-loop results found in \cite{Drukker:2011za}.

%%%%%%%%%%%%%%%%%%%%%%%%%%%%%%%%%%%%%%%%%%%%%%%%%%%%%%%%%%%%%%%%%%%%%%%%%%%%%%%%%%%%%%%%%%%%%%%%%%%%%%%%%%%%%%
%%%%%%%%%%%%%%%%%%%%%%%%%%%%%%%%%%%%%%%%%%%%%%%%%%%%%%%%%%%%%%%%%%%%%%%%%%%%%%%%%%%%%%%%%%%%%%%%%%%%%%%%%%%%%%
%%%%%%%%%%%%%%%%%%%%%%%%%%%%%%%%%%%%%%%%%%%%%%%%%%%%%%%%%%%%%%%%%%%%%%%%%%%%%%%%%%%%%%%%%%%%%%%%%%%%%%%%%%%%%%

\subsection{String ending on a line with an internal cusp}
\label{internal}

%%%%%%%%%%%%%%%%%%%%%%%%%%%%%%%%%%%%%%%%%%%%%%%%%%%%%%%%%%%%%%%%%%%%%%%%%%%%%%%%%%%%%%%%%%%%%%%%%%%%%%%%%%%%%%
%%%%%%%%%%%%%%%%%%%%%%%%%%%%%%%%%%%%%%%%%%%%%%%%%%%%%%%%%%%%%%%%%%%%%%%%%%%%%%%%%%%%%%%%%%%%%%%%%%%%%%%%%%%%%%
%%%%%%%%%%%%%%%%%%%%%%%%%%%%%%%%%%%%%%%%%%%%%%%%%%%%%%%%%%%%%%%%%%%%%%%%%%%%%%%%%%%%%%%%%%%%%%%%%%%%%%%%%%%%%%

In this limit, the induced geometry is again (\ref{inducedmetric}) but the mass terms in the quadratic operators are different. The fermionic fluctuation equations  are of the form
\be
\left(i\left(\partial_{\sigma}+\tfrac{{\rm sn}(\sigma){\rm dn}(\sigma)}{2 {\rm cn}(\sigma)}\right)\gamma^1
+\omega_n\gamma^0-m_F(\sigma)\frac{\sqrt{1-k^2}}{{\rm cn}(\sigma)}\right)\Psi_n = 0\,.
\label{gfermopint}
\ee
where now
\be
m_F(\sigma)=\frac{1}{4}\left(s_1-s_2+\frac{\textrm{dn}(\sigma)}{\sqrt{1-k^2}}s_3\right) +\frac{3}{4}\frac{\textrm{dn}(\sigma)}{\sqrt{1-k^2}}s_1 s_2 s_3\,.
\label{massesinternal}
\ee
Three of the scalar modes satisfy the equation
\be
\left(\partial^2_{\sigma}+\omega_n^2-\frac{m_B^2(1-k^2)}{\textrm{cn}^2(\sigma)}-k^2\right)\phi_n=0\,,
\label{intop}
\ee
two for $m_B^2=2$ and one for $m_B^2 = 0$. Four other scalar fields satisfy
\be
\left(\partial^2_{\sigma}+\omega_n^2-\frac{k^2}{4}\right)\phi_n=0\,,
\label{intop2}
\ee
while last scalar mode satisfies
\be
\left(\partial^2_{\sigma}+\omega_n^2+2k^2\textrm{cn}^2(\sigma)-k^2\right)\phi_n=0\,.
\label{intopR}
\ee

%%%%%%%%%%%%%%%%%%%%%%%%%%%%%%%%%%%%%%%%%%%%%%%%%%%%%%%%%%%%%%%%%%%%%%%%%%%%%%%%%%%%%%%%%%%%%%%%%%%%%%%%%%%%%%
%%%%%%%%%%%%%%%%%%%%%%%%%%%%%%%%%%%%%%%%%%%%%%%%%%%%%%%%%%%%%%%%%%%%%%%%%%%%%%%%%%%%%%%%%%%%%%%%%%%%%%%%%%%%%%
%%%%%%%%%%%%%%%%%%%%%%%%%%%%%%%%%%%%%%%%%%%%%%%%%%%%%%%%%%%%%%%%%%%%%%%%%%%%%%%%%%%%%%%%%%%%%%%%%%%%%%%%%%%%%%

\subsubsection{Small cusp}

%%%%%%%%%%%%%%%%%%%%%%%%%%%%%%%%%%%%%%%%%%%%%%%%%%%%%%%%%%%%%%%%%%%%%%%%%%%%%%%%%%%%%%%%%%%%%%%%%%%%%%%%%%%%%%
%%%%%%%%%%%%%%%%%%%%%%%%%%%%%%%%%%%%%%%%%%%%%%%%%%%%%%%%%%%%%%%%%%%%%%%%%%%%%%%%%%%%%%%%%%%%%%%%%%%%%%%%%%%%%%
%%%%%%%%%%%%%%%%%%%%%%%%%%%%%%%%%%%%%%%%%%%%%%%%%%%%%%%%%%%%%%%%%%%%%%%%%%%%%%%%%%%%%%%%%%%%%%%%%%%%%%%%%%%%%%

In the small $k^2$ limit eq. (\ref{intop}) becomes
\be
\left(\partial^2_{\tilde\sigma}+\tilde\omega_n^2-\frac{m_B^2}{\cos^2\tilde\sigma}
-k^2(1-\tfrac{m_B^2}{2})+{\cal O}(k^4)\right)\phi_n=0\,,
\label{intopexpa}
\ee
Thus, the two modes with $m_B^2 =2$ becomes unperturbed at this order with
\be
\omega_n = (n+2)\left(1-\frac{k^2}{4}\right)\,.
\ee
Their contribution to the vacuum energy is
\be
\frac12\sum_{n=0}^{\infty}\omega_n=\frac12\zeta_H(-1,2)\left(1-\frac{k^2}{4}\right)\,.
\label{intcont1}
\ee
The mode with $m_B^2 =0$ leads to
\be
\omega_n=(n+1)\left(1-\frac{k^2}{4}\right)+\frac{k^2}{2}\frac{1}{n+1}\,,
\ee
and summing these corrected frequencies we obtain the contribution to the vacuum energy
\be
\frac12\sum_{n=0}^{\infty}\omega_n=\frac12\zeta_H(-1,1)\left(1-\frac{k^2}{4}\right)+\frac{k^2}{4}\zeta_H(1,1)\,.
\label{intcont3}
\ee

Similarly, for the four modes satisfying (\ref{intop2})
\be
\omega_n=(n+1)\left(1-\frac{k^2}{4}\right)+\frac{k^2}{8}\frac{1}{n+1}\,,
\ee
each contributing
\be
\frac12\sum_{n=0}^{\infty}\omega_n=\frac12\zeta_H(-1,1)\left(1-\frac{k^2}{4}\right)+\frac{k^2}{16}\zeta_H(1,1)\,.
\label{intcont4}
\ee

The expansion of (\ref{intopR}) gives
\be
\left(\partial^2_{\tilde\sigma}+\tilde\omega_n^2 + k^2\cos 2\tilde{\sigma}+{\cal O}(k^2)\right)\phi_n=0\,,
\label{intopRexpa}
\ee
for which
\be
\tilde\omega_n^2 = (n+1)^2 -k^2\int_{-\tfrac{\pi}{2}}^{\tfrac{\pi}{2}}d\tilde{\sigma}\phi_{1,n}^* \cos^2\tilde{\sigma}\phi_{1,n}
=(n+1)^2 -\frac{k^2}{2} \delta_{n0}\,.
\ee
The contribution of this mode to the vacuum energy is then
\be
\frac{1}{2}\sum_{n=0}^\infty \omega_n =
\frac{1}{2}\zeta_H(-1,1)\left(1-\frac{k^2}{4}\right) -\frac{k^2}{8} +{\cal O}(k^4)\,.
\label{bcontR}
\ee

To obtain the contribution from the fermionic modes we evaluate the small $k$ expansion of (\ref{massesinternal}) for the 8 different combinations $(s_1,s_2,s_2)$. There are four modes with $m_F=\pm 1$ for which the perturbed equations in the rescaled variable is\footnote{Here and in what follows we
omit the perturbation from the spin connection because its contribution to the perturbed frequencies vanishes.}
\be
\left(i\left(\partial_{\tilde{\sigma}}+\tfrac{\tan\tilde{\sigma}}{2}\right)\gamma^1 +\tilde{\omega}_n\gamma^0-\frac{m_F}{\cos\tilde{\sigma}}-k^2\frac{m_F}{4}\cos\tilde{\sigma}\right)\Psi_n = 0\,.
\label{intfop1}
\ee
Their corrected frequencies are
\be
\omega_n=\left(n+\frac32\right)\left(1-\frac{k^2}{4}\right)+\frac{k^2}{4}\frac{1}{n+1}-\frac{k^2}{8}\frac{1}{(n+1)(n+2)}\,,
\label{omegak}
\ee
and so the sum over oscillator frequencies results
\be
\frac12\sum_{n=0}^{\infty}\omega_n=\frac{1}{2}\zeta_H(-1,\tfrac{3}{2})\left(1-\frac{k^2}{4}\right)+\frac{k^2}{8}\zeta_H(1,1)-\frac{k^2}{16}\,.
\label{intfcont2}
\ee

For the remaining two modes with $m_F=\pm 1$ the potential is vanishing
\be
\left(i\left(\partial_{\tilde{\sigma}}+\tfrac{\tan\tilde{\sigma}}{2}\right)\gamma^1 +\tilde{\omega}_n\gamma^0-\frac{m_F}{\cos\tilde{\sigma}}\right)\Psi_n = 0\,,
\label{intfop2}
\ee
and the frequencies are simply
\be
\omega_n=\left(n+\frac32\right)\left(1-\frac{k^2}{4}\right)\,,
\label{omegak2}
\ee
whose sum is
\be
\frac12\sum_{n=0}^{\infty}\omega_n=\frac{1}{2}\zeta_H(-1,\tfrac{3}{2})\left(1-\frac{k^2}{4}\right)\,.
\label{intfcont3}
\ee

For the two massless modes we obtain
\be
\left(i\left(\partial_{\tilde{\sigma}}+\tfrac{\tan\tilde{\sigma}}{2}\right)\gamma^1 +\tilde{\omega}_n\gamma^0 \pm \frac{k^2}{4}\cos\tilde{\sigma}\right)\Psi_n = 0\,,
\label{intfop0}
\ee
where the $\pm$ sign correspond to $(s_1 s_2 s_3)$ equal $(+-+)$ and $(-+-)$ respectively. At this point we have to decide which fermionic massless modes ($\Psi_{0,n}$ or ${\rm X}_{0,n}$) should be used to compute
the perturbative correction to the frequencies. As we have anticipated, only for a specific choice, the full set of fluctuations will conform an ${\cal N} =6$ supersymmetry multiplet. Moreover, since the background string configuration with cusp angles $\theta = \pm \phi$ is supersymmetric, among the four possible choices we select the one that leads to same cusp anomalous dimension as for the geometrical cusp angle. For that one has to take ${\rm X}_{0,n}$ for $(+-+)$ while $\Psi_{0,n}$ for $(-+-)$, which leads
for both modes to
\be
\omega_n=\left(n+\frac12\right)\left(1-\frac{k^2}{4}\right) -\frac{k^2}{8}\delta_{n0}\,,
\ee
and then
\be
%\omega_0=\frac12-\frac{k^2}{4}\rightarrow
\frac12\sum_{n=0}^{\infty}\omega_n=\frac12\zeta_H(-1,\tfrac{1}{2})\left(1-\frac{k^2}{4}\right)-\frac{k^2}{16}\,.
\label{intfcont1}
\ee

Now we sum all the contributions to the vacuum energy to $k^2$ order
\begin{align}
\frac{1}{T}\log{Z_{1-loop}}=& 4\left[\frac{1}{2}\zeta_H(-1,\tfrac{3}{2})\left(1-\frac{k^2}{4}\right) +\frac{k^2}{8}\zeta_H(1,1)-\frac{k^2}{16}\right]
+2 \left[\frac12\zeta_H(-1,\tfrac{1}{2})\left(1-\frac{k^2}{4}\right)-\frac{k^2}{16}\right] \nn\\
&+\frac22\zeta_H(-1,\tfrac12)\left(1-\frac{k^2}{4}\right)\nn\\
&-\frac{2}{2}\zeta_H(-1,2)\left(1-\frac{k^2}{4}\right) -\left[\frac{1}{2}\zeta_H(-1,1)\left(1-\frac{k^2}{4}\right)-\frac{k^2}{8}\right]\nn\\
&-\left[\frac12\zeta_H(-1,1)\left(1-\frac{k^2}{4}\right)+\frac{k^2}{4}\zeta_H(1,1)\right]-
4\left[\frac12\zeta_H(-1,1)\left(1-\frac{k^2}{4}\right)+\frac{k^2}{16}\zeta_H(1,1)\right]\nn\\
&=-\frac{k^2}{4}\,.
\end{align}
In this limit, $k^2=-\frac{\theta^2}{\pi^2}$. Therefore the cusp anomalous dimension for internal cusp results
\be
\Gamma^{cusp}_{1-loop}=-\frac{\theta^2}{4\pi^2}\,.
\label{gammaint}
\ee
This is consistent with the fact that cusped Wilson loop is BPS for $\theta=\pm\phi$ \cite{Griguolo:2012iq}. In other words, the Bremsstrahlung function up to 1-loop order in the
strong coupling limit is
\be
B^{\phi}=B^{\theta}=\frac{\sqrt{2\lambda \pi^2}}{4\pi^2} -\frac{1}{4\pi^2} + {\cal O}(\lambda^{-1/2})\,.
\label{final}
\ee

~

We can straightforwardly obtain the 1-loop partition function for a string with an internal cusp
in $AdS_5\times S^5$ from the previous computations. In that case one has eight $m_F=\pm1$ fermions contributing
with (\ref{intfcont2}), four massless scalars resulting in (\ref{intcont3}), three $m_B^2=2$ scalars giving (\ref{intcont1}) and one scalar
coupled to the worldsheet curvature contributing with (\ref{bcontR})
\begin{align}
\frac{1}{T}\log{Z_{1-loop}}=& 8\left[\frac{1}{2}\zeta_H(-1,\tfrac{3}{2})\left(1-\frac{k^2}{4}\right) +\frac{k^2}{8}\zeta_H(1,1)-\frac{k^2}{16}\right]
\nn\\
&-4\left[\frac12\zeta_H(-1,1)\left(1-\frac{k^2}{4}\right)+\frac{k^2}{4}\zeta_H(1,1)\right] -\frac{3}{2}\zeta_H(-1,2)\left(1-\frac{k^2}{4}\right)
\nn\\
& -\left[\frac{1}{2}\zeta_H(-1,1)\left(1-\frac{k^2}{4}\right)-\frac{k^2}{8}\right]
= -\frac{3k^2}{8}\,.
\end{align}
From $\log Z=-\Gamma_{\sf cusp}T$ and \eqref{BPS} we obtain
\be
\Gamma_{\rm cusp}^{1-loop}=-\frac{3\theta^2}{8\pi^2}\,, \qquad{\rm for}\qquad AdS_5 \times S^5\,,
\ee
in agreement with the 1-loop results found in \cite{Drukker:2011za}.

%%%%%%%%%%%%%%%%%%%%%%%%%%%%%%%%%%%%%%%%%%%%%%%%%%%%%%%%%%%%%%%%%%%%%%%%%%%%%%%%%%%%%%%%%%%%%%%%%%%%%%%%%%%%%%
%%%%%%%%%%%%%%%%%%%%%%%%%%%%%%%%%%%%%%%%%%%%%%%%%%%%%%%%%%%%%%%%%%%%%%%%%%%%%%%%%%%%%%%%%%%%%%%%%%%%%%%%%%%%%%
%%%%%%%%%%%%%%%%%%%%%%%%%%%%%%%%%%%%%%%%%%%%%%%%%%%%%%%%%%%%%%%%%%%%%%%%%%%%%%%%%%%%%%%%%%%%%%%%%%%%%%%%%%%%%%

\section{Exact Bremsstrahlung functions in ABJM}
\label{exact}

%%%%%%%%%%%%%%%%%%%%%%%%%%%%%%%%%%%%%%%%%%%%%%%%%%%%%%%%%%%%%%%%%%%%%%%%%%%%%%%%%%%%%%%%%%%%%%%%%%%%%%%%%%%%%%
%%%%%%%%%%%%%%%%%%%%%%%%%%%%%%%%%%%%%%%%%%%%%%%%%%%%%%%%%%%%%%%%%%%%%%%%%%%%%%%%%%%%%%%%%%%%%%%%%%%%%%%%%%%%%%
%%%%%%%%%%%%%%%%%%%%%%%%%%%%%%%%%%%%%%%%%%%%%%%%%%%%%%%%%%%%%%%%%%%%%%%%%%%%%%%%%%%%%%%%%%%%%%%%%%%%%%%%%%%%%%

The cusp anomalous dimension for small angles is given by the Bremsstrahlung function,
\be
\Gamma_{\rm cusp} = -B(\lambda,N)\phi^2 + {\cal O}(\phi^4)\,.
\ee
This Bremsstrahlung function has been related to the expectation value of certain BPS
circular Wilson loop, which is exactly known \cite{Lewkowycz:2013laa}
\be
B(\lambda,N) = \frac{1}{4\pi^2} \left.\partial_n \log|\langle W_n \rangle|\right|_{n=1}\,,
\label{mastereq2}
\ee
where $W_n$ is a Wilson loop that winds $n$ times around the maximal circle in a 3-sphere.

{In this section, we review the strong coupling expansion of the Bremsstrahlung functions
from equation (\ref{mastereq2}). The strong coupling expansion of the 1/6 BPS multiply wound circular
Wilson loop is \cite{Klemm:2012ii},
\begin{alignat}{3}
\langle W_n \rangle & =  \frac{i^ne^{\pi n \sqrt{2\lambda}}}{\lambda}
\left(\frac{\sqrt{2\lambda}}{4\pi n} -\frac{H_n}{4 \pi^2 n} - \frac{i}{8\pi n} -\frac{1}{96}
+ \left(\frac{i}{192}+\frac{\pi n}{4608}+ \frac{H_{n-1}}{96\pi}\right)\frac{1}{\sqrt{2\lambda}}
+{\cal O}(\lambda^{-1})\right),\nn
\\
\label{Wexpansionstrong}
\end{alignat}
where $H_n$ are harmonic numbers. The use of this expansion in (\ref{mastereq2}) gives the Bremsstrahlung function
for a small cusp deforming a 1/6 BPS Wilson line \cite{Lewkowycz:2013laa}
\be
B^{1/6}(\lambda)  =  \frac{\sqrt{2\lambda \pi^2}}{4\pi^2} -\frac{1}{4\pi^2} +
\left(\frac{1}{4\pi^2}-\frac{5}{96}\right)\frac{1}{\sqrt{2\lambda \pi^2}} + {\cal O}(\lambda^{-1})\,.
\label{B16}
\ee

To obtain the Bremsstrahlung function for a small cusp deforming a 1/2 BPS Wilson line, we
need to feed (\ref{mastereq2}) with the 1/2 BPS multiply wound circular Wilson loop expectation
value \cite{Klemm:2012ii}
\be
\langle W_n^{1/2} \rangle = \langle W_n\rangle - e^{i n\pi} \langle \overline{W}_n \rangle\,,
\ee
where $W_n$ is the previous 1/6 circular Wilson and  $\langle\overline{W}_n\rangle$ is the complex conjugate
of $\langle{W}_n\rangle$.
We then obtain
\be
B^{1/2}(\lambda) = \frac{1}{4\pi^2} \left.\partial_n \log (\langle W_n \rangle + \langle \overline{W}_n \rangle)\right|_{n=1}
-\frac{i}{8\pi}\frac{\langle W_1 \rangle -  \langle \overline{W}_1 \rangle}{\langle W_1 \rangle + \langle \overline{W}_1 \rangle}\,.
\ee
Since the derivative $\partial_n (\langle W_n \rangle + \langle \overline{W}_n \rangle)$ vanishes at $n=1$, we are left with
\be
B^{1/2}(\lambda) = \frac{1}{8\pi} \frac{ {\rm Im}\left({\langle W_1 \rangle}\right)}{ {\rm Re }\left({\langle W_1 \rangle}\right) }\,.
\ee
Using the expansion (\ref{Wexpansionstrong}), one obtains \cite{Bianchi:2014laa}
\begin{alignat}{2}
B^{1/2}(\lambda) & =  \frac{\sqrt{2\lambda \pi^2}}{4\pi^2} -\frac{1}{4\pi^2}-
\frac{1}{96}\frac{1}{\sqrt{2\lambda \pi^2}} + {\cal O}(\lambda^{-1})\,.
\label{Bexpansions}
\end{alignat}
The first two orders of \nref{Bexpansions} are in complete agreement with \nref{final}, computed
in section \ref{anomaly}.
}

%%%%%%%%%%%%%%%%%%%%%%%%%%%%%%%%%%%%%%%%%%%%%%%%%%%%%%%%%%%%%%%%%%%%%%%%%%%%%%%%%%%%%%%%%%%%%%%%%%%%%%%%%%%%%%
%%%%%%%%%%%%%%%%%%%%%%%%%%%%%%%%%%%%%%%%%%%%%%%%%%%%%%%%%%%%%%%%%%%%%%%%%%%%%%%%%%%%%%%%%%%%%%%%%%%%%%%%%%%%%%
%%%%%%%%%%%%%%%%%%%%%%%%%%%%%%%%%%%%%%%%%%%%%%%%%%%%%%%%%%%%%%%%%%%%%%%%%%%%%%%%%%%%%%%%%%%%%%%%%%%%%%%%%%%%%%
\section{Discussion}
\label{discussion}
%%%%%%%%%%%%%%%%%%%%%%%%%%%%%%%%%%%%%%%%%%%%%%%%%%%%%%%%%%%%%%%%%%%%%%%%%%%%%%%%%%%%%%%%%%%%%%%%%%%%%%%%%%%%%%
%%%%%%%%%%%%%%%%%%%%%%%%%%%%%%%%%%%%%%%%%%%%%%%%%%%%%%%%%%%%%%%%%%%%%%%%%%%%%%%%%%%%%%%%%%%%%%%%%%%%%%%%%%%%%%
%%%%%%%%%%%%%%%%%%%%%%%%%%%%%%%%%%%%%%%%%%%%%%%%%%%%%%%%%%%%%%%%%%%%%%%%%%%%%%%%%%%%%%%%%%%%%%%%%%%%%%%%%%%%%%

We have computed the partition function for strings ending in cusped lines at the boundary to 1-loop order in
the large $\sqrt{\lambda}$ expansion. We computed the corresponding 1-loop determinants by relating them
to the vacuum energy density of the 1-loop oscillator modes.

These 1-loop determinants have been computed previously for small cusp angles, by reducing the operators to
1-dimensional ordinary differential ones and exploiting the Gelfand-Yaglom method \cite{Forini:2010ek,Beccaria:2010ry,Drukker:2011za,forini}.
In the case of strings in $AdS_5\times S^5$, our results coincide with those of \cite{Drukker:2011za}.
For strings in $AdS_4\times \mathbb{CP}^3$, our results differ from those in \cite{forini}.
We understand the discrepancy in the result as due to the imposition of different boundary conditions for
the massless modes.

In two dimensions the Dirac operator can be reduced to the flat space form by rescaling
the spinor with $g^{1/4}$. From for the $AdS_2$ fermionic modes \eqref{feigen} one obtains, after the
rescaling, the following spinor solution to the flat space massless Dirac operator
\be
\begin{array}{c}
\tilde\psi^1_{0,n}(\sigma)\equiv g^{1/4}\psi^1_{0,n}(\sigma)=
\frac{1}{\sqrt\pi}\cos\left((n+\tfrac{1}{2})\sigma-(n-\tfrac{1}{2})\frac{\pi}{2}\right)
\\
\tilde\psi^2_{0,n}(\sigma) \equiv g^{1/4}\psi^2_{0,n}(\sigma)=
\frac{1}{\sqrt\pi}\cos\left((n+\tfrac{1}{2})\sigma-(n+\tfrac{5}{2})\frac{\pi}{2}\right)
\end{array}\,.
\ee
Each  component of these solutions with $\omega=n+\tfrac{1}{2}$ satisfies Dirichlet boundary conditions in one endpoint and Neumann boundary conditions in the other. Other solutions with $\omega = n$ exist. They have Dirichlet boundary conditions
for one component and Neumann boundary conditions for the other. However, they do not
form a supersymmetry multiplet with the massless scalar fields we are using.

When using the off-shell method to compute 1-loop determinants, the choice of boundary conditions
will impact in the outcome as well. In \cite{forini} Dirichlet boundary conditions were used for
the fermionic fluctuations. The determinant of the corresponding quadratic operator with spectrum $\lambda_n = {\tilde\omega}^2 + n^2$
could then be simplified with the determinant of a massless scalar field with Dirichlet boundary conditions.

Let us briefly analyze how the results of \cite{forini} would change if the determinant of the quadratic operator
for the massless fermionic modes were computed with Dirichlet-Neumann boundary conditions. The corresponding
spectrum would be $\lambda_n = {\tilde\omega}^2 + (n+\tfrac{1}{2})^2$. Using that
\be
\frac{\det(\partial_{\tilde\sigma}^2-\tilde\omega^2)_{DN}}{\det(\partial_{\tilde\sigma}^2-\tilde\omega^2)_{DD}} = {\tilde\omega}\coth(\tilde\omega\pi)\,,
\ee
it is straightforward to see that the resulting change in the geometrical cusp anomalous dimension is
\begin{alignat}{2}
\delta\Gamma_{\phi} =& -\frac{T}{4\pi}\int_{-\infty}^{\infty}d\omega\log(\coth^2(2 K \omega))
= - \frac{T \pi}{16 K} \nn\\
=& -T\left(\frac{1}{8}-\frac{k^2}{32}+{\cal O}(k^4)\right)\,.
\label{correction}
\end{alignat}
The result computed in \cite{forini} corrected by (\ref{correction}) is in complete agreement
with our result.

Our results for the cusp anomalous dimension for a geometrical cusp in section \ref{geometrical}
and for an internal cusp in section \ref{internal} are consistent with its vanishing in the BPS case
$\theta = \pm\phi$. Moreover this unique Bremsstrahlung function found in the strong coupling limit
(\ref{final}) is in perfect agreement with the expansion of the exact result (\ref{Bexpansions})
in section \ref{exact}.

Finally, we would also like to comment about the Bremsstrahlung function for a small geometrical cusp
deformation of a 1/6 BPS Wilson line. From the string point of view, this kind of Wilson loops are interpreted as the smearing of string
configurations over a $\mathbb {CP}^1\subset  \mathbb {CP}^3$. This corresponds
to impose Neumann boundary conditions for the directions along the $ \mathbb {CP}^1$ \cite{Lewkowycz:2013laa}.

Therefore, to compute $Z_{1-loop}$ for geometrical cusp angle deformation of a 1/6 BPS Wilson line,
we should repeat the perturbative analysis replacing two of the scalar massless modes with Dirichlet
boundary conditions by two scalar massless modes with Neumann boundary conditions, and correspondingly replacing
fermionic massless modes type $\Psi_{0,n}$ by ${\rm X}_{0,n}$ and viceversa.
However, in this case, the perturbation of scalar massless modes vanishes using either (\ref{beigen}) or
(\ref{neumann}). The same happens for fermionic massless modes and, then, to this order
$Z_{1-loop}$ is the same as for the deformation of a 1/2 BPS Wilson line
\be
B^{\phi}_{1/6}=\frac{\sqrt{2\lambda \pi^2}}{4\pi^2} -\frac{1}{4\pi^2}
+ {\cal O}(\lambda^{-1/2})\,,
\label{final2}
\ee
which is in agreement with (\ref{B16}). For an internal cusp angle, the dual string has a non-trivial profile
inside the $ \mathbb {CP}^3$. The smearing of such configurations has no  clear geometrical interpretation \cite{Correa:2014aga}.

~

%%%%%%%%%%%%%%%%%%%%%%%%%%%%%%%%%%%%%%%%%%%%%%%%%%%%%%%%%%%%%%%%%%%%%%%%%%%%%%%%%%%%%%%%%%%%%%%%%%%%%%%%%%%%%%
%%%%%%%%%%%%%%%%%%%%%%%%%%%%%%%%%%%%%%%%%%%%%%%%%%%%%%%%%%%%%%%%%%%%%%%%%%%%%%%%%%%%%%%%%%%%%%%%%%%%%%%%%%%%%%
%%%%%%%%%%%%%%%%%%%%%%%%%%%%%%%%%%%%%%%%%%%%%%%%%%%%%%%%%%%%%%%%%%%%%%%%%%%%%%%%%%%%%%%%%%%%%%%%%%%%%%%%%%%%%%

{\bf Acknowledgements }

We would like to thank V.Forini, V.Giangreco M. Pulleti and O.Ohlsson Sax for useful comments.
This work was supported by CONICET and grants PICT 2012-0417, PIP 0681 and PIP 0595/13.

%%%%%%%%%%%%%%%%%%%%%%%%%%%%%%%%%%%%%%%%%%%%%%%%%%%%%%%%%%%%%%%%%%%%%%%%%%%%%%%%%%%%%%%%%%%%%%%%%%%%%%%%%%%%%%
%%%%%%%%%%%%%%%%%%%%%%%%%%%%%%%%%%%%%%%%%%%%%%%%%%%%%%%%%%%%%%%%%%%%%%%%%%%%%%%%%%%%%%%%%%%%%%%%%%%%%%%%%%%%%%
%%%%%%%%%%%%%%%%%%%%%%%%%%%%%%%%%%%%%%%%%%%%%%%%%%%%%%%%%%%%%%%%%%%%%%%%%%%%%%%%%%%%%%%%%%%%%%%%%%%%%%%%%%%%%%

%%%%%%%%%%%%%%%%%%%%%%%%%%%%%%%%%%%%%%%%%%%%%%%%%%%%%%%%%%%%%%%%%%%%%%%%%%%%%%%%%%%%%%%%%%%%%%%%%%%%%%%%%%%%%%%%%%%%%%%%%%%%%%%%%%%%%%%%%%%%%%%%%%%%%%%%%%%%%%%%%%%%%%%%%%%%%%%%%%%%
\appendix
%%%%%%%%%%%%%%%%%%%%%%%%%%%%%%%%%%%%%%%%%%%%%%%%%%%%%%%%%%%%%%%%%%%%%%%%%%%%%%%%%%%%%%%%%%%%%%%%%%%%%%%%%%%%%%%%%%%%%%%%%%%%%%%%%%%%%%%%%%%%%%%%%%%%%%%%%%%%%%%%%%%%%%%%%%%%%%%%%%%%

%%%%%%%%%%%%%%%%%%%%%%%%%%%%%%%%%%%%%%%%%%%%%%%%%%%%%%%%%%%%%%%%%%%%%%%%%%%%%%%%%%%%%%%%%%%%%%%%%%%%%%%%%%%%%%
%%%%%%%%%%%%%%%%%%%%%%%%%%%%%%%%%%%%%%%%%%%%%%%%%%%%%%%%%%%%%%%%%%%%%%%%%%%%%%%%%%%%%%%%%%%%%%%%%%%%%%%%%%%%%%
%%%%%%%%%%%%%%%%%%%%%%%%%%%%%%%%%%%%%%%%%%%%%%%%%%%%%%%%%%%%%%%%%%%%%%%%%%%%%%%%%%%%%%%%%%%%%%%%%%%%%%%%%%%%%%

\section{Review of the classical solution and quantum fluctuations}
\label{classical}

%%%%%%%%%%%%%%%%%%%%%%%%%%%%%%%%%%%%%%%%%%%%%%%%%%%%%%%%%%%%%%%%%%%%%%%%%%%%%%%%%%%%%%%%%%%%%%%%%%%%%%%%%%%%%%
%%%%%%%%%%%%%%%%%%%%%%%%%%%%%%%%%%%%%%%%%%%%%%%%%%%%%%%%%%%%%%%%%%%%%%%%%%%%%%%%%%%%%%%%%%%%%%%%%%%%%%%%%%%%%%
%%%%%%%%%%%%%%%%%%%%%%%%%%%%%%%%%%%%%%%%%%%%%%%%%%%%%%%%%%%%%%%%%%%%%%%%%%%%%%%%%%%%%%%%%%%%%%%%%%%%%%%%%%%%%%

In this section we review the classical solution and its quantum fluctuations found in \cite{forini} for a string in $AdS_4\times\mathbb{CP}^3$ ending on a Wilson line with geometrical cusp $\phi$ and internal cusp $\theta$.

%%%%%%%%%%%%%%%%%%%%%%%%%%%%%%%%%%%%%%%%%%%%%%%%%%%%%%%%%%%%%%%%%%%%%%%%%%%%%%%%%%%%%%%%%%%%%%%%%%%%%%%%%%%%%%
%%%%%%%%%%%%%%%%%%%%%%%%%%%%%%%%%%%%%%%%%%%%%%%%%%%%%%%%%%%%%%%%%%%%%%%%%%%%%%%%%%%%%%%%%%%%%%%%%%%%%%%%%%%%%%
%%%%%%%%%%%%%%%%%%%%%%%%%%%%%%%%%%%%%%%%%%%%%%%%%%%%%%%%%%%%%%%%%%%%%%%%%%%%%%%%%%%%%%%%%%%%%%%%%%%%%%%%%%%%%%

\subsection{Classical solution}

%%%%%%%%%%%%%%%%%%%%%%%%%%%%%%%%%%%%%%%%%%%%%%%%%%%%%%%%%%%%%%%%%%%%%%%%%%%%%%%%%%%%%%%%%%%%%%%%%%%%%%%%%%%%%%
%%%%%%%%%%%%%%%%%%%%%%%%%%%%%%%%%%%%%%%%%%%%%%%%%%%%%%%%%%%%%%%%%%%%%%%%%%%%%%%%%%%%%%%%%%%%%%%%%%%%%%%%%%%%%%
%%%%%%%%%%%%%%%%%%%%%%%%%%%%%%%%%%%%%%%%%%%%%%%%%%%%%%%%%%%%%%%%%%%%%%%%%%%%%%%%%%%%%%%%%%%%%%%%%%%%%%%%%%%%%%

The target space is $AdS_4\times\mathbb{CP}^3$ with metric
\be
ds^2=R^2\left(ds^2_{AdS_4}+4ds^2_{\mathbb{CP}^3}\right)\,,
\ee
where
\be
ds^2_{AdS_4}=-\cosh^2 \rho dt^2 + d\rho^2 + \sinh^2\rho\left(d\psi^2+\sin^2\psi d\varphi^2\right)\,.
\label{ads}
\ee
In this background a Green-Schwartz action (GS) is considered. The action of bosonic degrees of freedom
is of the Nambu-Goto type in the static gauge, taking $t$ and $\varphi$ as the worldsheet coordinates,
\be
S=\sqrt{\frac{\lambda}{2}}\int d^2\sigma \sqrt{\det G_{\mu\nu}\partial_iX^{\mu}\partial_jX^{\nu}}\,,
\ee
where $\mu,\nu$ and $i,j$ are target space and worldsheet indices respectively, $G_{\mu\nu}$ is the target space metric and $X^{\mu}$ are the embedding coordinates.

The global radius $\rho$ and $\vartheta$, a particular Killing direction in $\mathbb{CP}^3$,
are taken to be functions of $\varphi$
\be
\rho=\rho(\varphi)\,,\qquad \vartheta=\vartheta(\varphi)\,,\qquad
\label{ansatz}
\ee
while the remaining variables are held fixed. The cusp angles $\phi$ and $\theta$ relate to the angular extension of the string in $AdS$ and $\mathbb{CP}^3$ respectively as $\varphi\in\left[\phi/2,\pi-\phi/2\right]$ and  $\vartheta\in\left[-\theta/2,\theta/2\right]$. The conserved charges associated to $t$ and $\vartheta$ translations are
\be
E=-\frac{\sinh^2\rho\cosh\rho}{\sqrt{\sinh^2\rho+(\partial_{\varphi}\rho)^2+(\partial_{\varphi}\vartheta)^2}}\,,\qquad J=\frac{\partial_{\varphi}\vartheta\cosh\rho}{\sqrt{\sinh^2\rho+(\partial_{\varphi}\rho)^2+(\partial_{\varphi}\vartheta)^2}}\,.
\label{conserved}
\ee
The BPS condition $\theta=\pm\phi$ is equivalent to $E=\pm J$ at the classical level. Now we introduce two parameters
\be
p=\frac{1}{E}\,,\qquad q=-\frac{J}{E}\,,
\ee
in terms of which we can express the cusp angles
\be
\phi=\pi-2\frac{p^2}{b\sqrt{b^4+p^2}}\left[\Pi\left(\frac{b^4}{b^4+p^2}\mid k^2\right)-\mathbb{K}(k^2)\right]\,,
\label{geomcusp}
\qquad\theta=\frac{2bq}{\sqrt{b^4+p^2}}\mathbb{K}(k^2)\,,
\ee
where
\be
p^2=\frac{b^4(1-k^2)}{b^2+k^2}\,,\qquad q^2=\frac{b^2(1-2k^2-k^2b^2)}{b^2+k^2}\,.
\ee
The small cusp angles limit ($\phi,\theta\ll 1$) corresponds to $p\to\infty$
\be
\phi=\frac{\pi}{p}+\frac{\pi(3q^2-5)}{4p^3}+\mathcal{O}(p^{-5})\,,\qquad
\theta=\frac{\pi q}{p}+\frac{\pi q(q^2-3)}{4p^3}+\mathcal{O}(p^{-5})\,.
\label{intexp}
\ee
The classical action for this embedding reads
\be
S_{cl}=T\sqrt{2\lambda}\frac{b^4+p^2}{bp}\left[\frac{(b^2+1)p^2}{b^4+p^2}\mathbb{K}(k^2)-\mathbb{E}(k^2)\right]\,,
\ee
where $T$ is a cut-off for the time integration and a divergent term from the $\rho\to\infty$ limit
has been dropped.

%%%%%%%%%%%%%%%%%%%%%%%%%%%%%%%%%%%%%%%%%%%%%%%%%%%%%%%%%%%%%%%%%%%%%%%%%%%%%%%%%%%%%%%%%%%%%%%%%%%%%%%%%%%%%%
%%%%%%%%%%%%%%%%%%%%%%%%%%%%%%%%%%%%%%%%%%%%%%%%%%%%%%%%%%%%%%%%%%%%%%%%%%%%%%%%%%%%%%%%%%%%%%%%%%%%%%%%%%%%%%
%%%%%%%%%%%%%%%%%%%%%%%%%%%%%%%%%%%%%%%%%%%%%%%%%%%%%%%%%%%%%%%%%%%%%%%%%%%%%%%%%%%%%%%%%%%%%%%%%%%%%%%%%%%%%%

\subsection{Fluctuation Lagrangian}

%%%%%%%%%%%%%%%%%%%%%%%%%%%%%%%%%%%%%%%%%%%%%%%%%%%%%%%%%%%%%%%%%%%%%%%%%%%%%%%%%%%%%%%%%%%%%%%%%%%%%%%%%%%%%%
%%%%%%%%%%%%%%%%%%%%%%%%%%%%%%%%%%%%%%%%%%%%%%%%%%%%%%%%%%%%%%%%%%%%%%%%%%%%%%%%%%%%%%%%%%%%%%%%%%%%%%%%%%%%%%
%%%%%%%%%%%%%%%%%%%%%%%%%%%%%%%%%%%%%%%%%%%%%%%%%%%%%%%%%%%%%%%%%%%%%%%%%%%%%%%%%%%%%%%%%%%%%%%%%%%%%%%%%%%%%%
It is convenient to define a world-sheet coordinates $\tau$ and $\sigma$ for which
\be
\cosh^2\rho=\frac{1+b^2}{b^2\cn^2(\sigma\mid k^2)}\,,
\ee
with ranges
\be
-\mathbb{K}(k^2)<\sigma<\mathbb{K}(k^2)\,,\qquad\qquad -\infty<\tau<\infty\,.
\label{range}
\ee
In these coordinates, the induced metric takes the form
\be
ds^2=\frac{1-k^2}{\cn^2(\sigma\mid k^2)}\left(-d\tau^2+d\sigma^2\right)\,,
\label{conformal}
\ee
and leads to the following scalar curvature
\be
R^{(2)}=-2\left(1+\frac{k^2}{1-k^2}\cn^4(\sigma\mid k^2)\right)\,.
\ee

In what follows we drop the $k^2$ dependence in the Jacobi functions in order to simplify the notation.

Since we are in static gauge, there are no fluctuations in the longitudinal directions. Defining appropriate scalar fluctuation fields $\zeta_a$ for the transverse directions ($a=1,\ldots,8$) (see \cite{forini}) the bosonic quadratic action results
\be
S_b=\frac12\int d\tau d\sigma\sqrt{g}\left[g^{ij}\partial_i\zeta_{a}\partial_j\zeta_{b}+A\left(\zeta_8\partial_{\sigma}\zeta_7-\zeta_7\partial_{\sigma}\zeta_8\right)+M_{ab}\zeta_a\zeta_b\right]\,,
\ee
where
\begin{alignat}{2}
M_{11}&=\frac{b^4-b^2p^2-p^4}{b^2p^2\cosh^2\rho}+2\,,\qquad M_{22}=\frac{b^4-b^2p^2-p^4}{b^2p^2\cosh^2\rho}\,,
\\
M_{ss}&=\frac{b^4-b^2p^2-p^4}{4b^2p^2\cosh^2\rho}\,, \qquad \ s=3,4,5,6
\\
M_{77}&=\frac{b^4-b^2p^2-p^4}{b^2p^2\cosh^2\rho}-\frac{2(b^2+1)(b^2-p^2)}{b^2p^2\cosh^4\rho}+b^2\frac{b^4+2b^2p^2\sinh^2\rho+b^2p^2-p^2}{\cosh^2\rho(b^4+2b^2p^2\sinh^2\rho-p^2)^2}\,,
\\
M_{88}&=\frac{b^4-b^2p^2-p^4}{b^2p^2\cosh^2\rho}+2-\frac{3b^2}{\cosh^2\rho(b^4+2b^2p^2\sinh^2\rho-p^2)}+\frac{b^4p^2}{(b^4+2b^2p^2\sinh^2\rho-p^2)^2}\,,
\\
M_{78}&=M_{87}=\frac{2\sqrt{-b^4+b^2p^2+p^2}\sqrt{b^2\sinh^2\rho-1}\sqrt{b^2+p^2\sinh^2\rho}}{p\cosh^3\rho(b^4+2b^2p^2\sinh^2\rho-p^2)}\,,
\\
A&=\frac{2\sqrt{b^4+p^2}\sqrt{-b^4+b^2p^2+p^2}}{p\cosh^2\rho(b^4+2b^2p^2\sinh^2\rho-p^2)}\,.
\end{alignat}

For the fermionic modes the resulting quadratic fluctuation Lagrangian is of the Dirac type.
We will only write the expressions in the corresponding limits of small geometrical and internal cusp in the next subsection. The kappa symmetry fixing condition is of the form
\be
\frac12\left(1+\Gamma_{01}\Gamma_{11}\right)\theta=\theta\,.
\label{kappa}
\ee

%%%%%%%%%%%%%%%%%%%%%%%%%%%%%%%%%%%%%%%%%%%%%%%%%%%%%%%%%%%%%%%%%%%%%%%%%%%%%%%%%%%%%%%%%%%%%%%%%%%%%%%%%%%%%%
%%%%%%%%%%%%%%%%%%%%%%%%%%%%%%%%%%%%%%%%%%%%%%%%%%%%%%%%%%%%%%%%%%%%%%%%%%%%%%%%%%%%%%%%%%%%%%%%%%%%%%%%%%%%%%
%%%%%%%%%%%%%%%%%%%%%%%%%%%%%%%%%%%%%%%%%%%%%%%%%%%%%%%%%%%%%%%%%%%%%%%%%%%%%%%%%%%%%%%%%%%%%%%%%%%%%%%%%%%%%%

\subsubsection{Small cusp limit}

%%%%%%%%%%%%%%%%%%%%%%%%%%%%%%%%%%%%%%%%%%%%%%%%%%%%%%%%%%%%%%%%%%%%%%%%%%%%%%%%%%%%%%%%%%%%%%%%%%%%%%%%%%%%%%
%%%%%%%%%%%%%%%%%%%%%%%%%%%%%%%%%%%%%%%%%%%%%%%%%%%%%%%%%%%%%%%%%%%%%%%%%%%%%%%%%%%%%%%%%%%%%%%%%%%%%%%%%%%%%%
%%%%%%%%%%%%%%%%%%%%%%%%%%%%%%%%%%%%%%%%%%%%%%%%%%%%%%%%%%%%%%%%%%%%%%%%%%%%%%%%%%%%%%%%%%%%%%%%%%%%%%%%%%%%%%

We are interested in the behaviour of the 1-loop partition function when the cusp angle, either geometrical or internal, is small.

In the first case we consider a small geometrical cusp angle $\phi$ and a vanishing $\theta$. From \eqref{intexp} we conclude that $\theta=0$ implies $q\to 0$. In this limit we can express $p$ and $b$ in terms of $k$. For small values of $k$ the geometrical cusp angle is also small, $\phi^2 = \pi^2 k^2 + {\cal O}(k^4)$.

For bosonic modes the mass matrix become diagonal and we obtain the corresponding Klein-Gordon operators in the metric \eqref{conformal} with masses
\be
M_{11}=2 ,\qquad M_{88}=R^{(2)}+4 , \qquad M_{ss}=0 , \quad s=2,3,4,5,6,7
\label{geommass}
\ee

For fermionic modes we obtain a Dirac Lagrangian with covariant derivatives in metric \eqref{conformal} and mass term of the form
\be
M_F=\frac{i\Gamma_{01}}{4}\left[\left(\Gamma_{49}-\Gamma_{57}+\Gamma_{68}\right)-3\Gamma_{23}\right]\,,
\label{geomfermmass}
\ee
Now we can expand the 10-dimensional spinor in a basis in which matrices $\{i\Gamma_{49},i\Gamma_{57},i\Gamma_{68},i\Gamma_{23}\}$
are diagonal with eigenvalues $\{s_1,s_2,s_3,s_4\}$ ($s_i=\pm1$). With this choice that the Lagrangian factorizes into eight Lagrangians for 2-dimensional spinors, for which ${\Gamma_{01}\to\tilde{\gamma}_*}$  with {$\tilde{\gamma}_*$} the chiral gamma matrix. The kappa symmetry fixing \eqref{kappa} is equivalent to $s_4=-s_1s_2s_3$  and evaluating \eqref{geomfermmass} for the eight combinations of $\{s_1,s_2,s_3\}$ give two massless modes, three fermions with chiral mass $m_F=1$ and three fermions with $m_F=-1$.

The resulting 1-loop partition function then results
\be
Z_{1-loop}=\frac{\det^{2/2}\left(i{\tilde\gamma}^aD_a\right)\det^{3/2}\left(i\tilde\gamma^aD_a
-{\tilde{\gamma}_*}\right)
\det^{3/2}\left(i\tilde\gamma^aD_a+{ \tilde{\gamma}_*}\right)}
{\det^{6/2}\left(-\nabla^2\right)\det^{1/2}\left(-\nabla^2+R^{(2)}+4\right)\det^{1/2}\left(-\nabla^2+2\right)}\,,
\label{geomZ}
\ee
where $\tilde\gamma^a$ are curved 2-dimensional Dirac matrices for metric \eqref{conformal}, $D_a$ are the corresponding covariant derivatives namely
\be
D_{\tau}=\partial_{\tau}+\frac{\sn(\sigma)\dn(\sigma)}{2\cn(\sigma)}
%\sigma^2\sigma^1
{\gamma^0\gamma^1},\qquad D_{\sigma}=\partial_{\sigma}\,,
\ee
and $\nabla^2$ is the scalar Laplacian
\be
\nabla^2=\frac{\cn^2(\sigma)}{1-k^2}\left(\partial^2_{\tau}+\partial^2_{\sigma}\right)\,.
\ee
Note that in the $k\to 0$ limit they reduce to $AdS_2$ Dirac and Klein-Gordon operators and \eqref{geomZ} reduces to the partition function for a string ending on a straight line.
%the no cusped case \cite{DGT}.

The other case we study is a small internal cusp $\theta$ with vanishing $\phi$. This corresponds to $p\to\infty$ limit with $q/p$ and $b/p$ held fixed, namely
\be
\frac{q}{p}=\frac{ik}{\sqrt{1-k^2}}, \qquad \frac{b}{p}=\frac{1}{\sqrt{1-k^2}}\,.
\ee
Note that, in this limit, $k$ will be again the small parameter for the small $\theta$ expansion but now $k$ is imaginary. In particular, we have from \eqref{intexp} that, up to $k^2$ order, $k^2=-\theta^2/\pi^2$.

For the bosonic modes we obtain Klein-Gordon operators in metric \eqref{conformal} with mass terms
\be
\begin{array}{ll}
M_{11}=M_{88}=2+\frac{k^2}{\sqrt{g}}\,, & \quad M_{22}=\frac{k^2}{\sqrt{g}}\,, \\
M_{77}=R^{(2)}+2+\frac{k^2}{\sqrt{g}}\,, & \quad M_{ss}=\frac{k^2}{4\sqrt{g}}, \quad s=3,4,5,6
\end{array}
\label{intmass}
\ee

For fermionic modes the mass term in the 10d Dirac Lagrangian results
\be
M_F=\frac{i\Gamma_{01}}{4}\left[\left(\Gamma_{68}-\Gamma_{57}+\frac{\textrm{dn}(\sigma)}{\sqrt{1-k^2}}\Gamma_{49}\right)-3\frac{\textrm{dn}(\sigma)}{\sqrt{1-k^2}}\Gamma_{23}\right]\,.
\label{intfermmass}
\ee
Again we can expand the 10d spinor in the basis we used for the geometrical cusp obtaining the mass spectrum involved in the computation of Section \ref{internal}.

%%%%%%%%%%%%%%%%%%%%%%%%%%%%%%%%%%%%%%%%%%%%%%%%%%%%%%%%%%%%%%%%%%%%%%%%%%%%%%%%%%%%%%%%%%%%%%%%%%%%%%%%%%%%%%
%%%%%%%%%%%%%%%%%%%%%%%%%%%%%%%%%%%%%%%%%%%%%%%%%%%%%%%%%%%%%%%%%%%%%%%%%%%%%%%%%%%%%%%%%%%%%%%%%%%%%%%%%%%%%%
%%%%%%%%%%%%%%%%%%%%%%%%%%%%%%%%%%%%%%%%%%%%%%%%%%%%%%%%%%%%%%%%%%%%%%%%%%%%%%%%%%%%%%%%%%%%%%%%%%%%%%%%%%%%%%

\section{Rescaling of 1-loop operators}\label{rescaling}

%%%%%%%%%%%%%%%%%%%%%%%%%%%%%%%%%%%%%%%%%%%%%%%%%%%%%%%%%%%%%%%%%%%%%%%%%%%%%%%%%%%%%%%%%%%%%%%%%%%%%%%%%%%%%%
%%%%%%%%%%%%%%%%%%%%%%%%%%%%%%%%%%%%%%%%%%%%%%%%%%%%%%%%%%%%%%%%%%%%%%%%%%%%%%%%%%%%%%%%%%%%%%%%%%%%%%%%%%%%%%
%%%%%%%%%%%%%%%%%%%%%%%%%%%%%%%%%%%%%%%%%%%%%%%%%%%%%%%%%%%%%%%%%%%%%%%%%%%%%%%%%%%%%%%%%%%%%%%%%%%%%%%%%%%%%%

As explained in \cite{DGT}, for certain Wilson loops in $AdS_5\times S^5$ type IIB string theory, we can
compute the 1-loop correction to the partition function in terms of the vacuum energy, {\it i.e.} the sum over the oscillator modes of
the 1-loop fluctuation operators . We will show that this statement is also valid for the Wilson loops discussed in this work.  For this result to be correct we need to rescale the quadratic operators by $\mathcal{M}=g^{00}$ \cite{Camporesi:1992wn}. This can be made by a redefinition of the
internal product, which implies a Weyl rescaling of the metric. At the quantum level, a conformal rescaling of the metric introduces a conformal anomaly, which depends on the fields involved and the equations of motion they satisfy, \cite{Schwarz:1979ae,Alvarez:1982zi,Luckock:1989mv}.

Consider a general second order operator acting on scalar fields in the following way
\be
\left(\phi,\Delta^B\phi\right)=\int d^2\sigma\sqrt{g}\left(g^{ij}\partial_i\phi\partial_j\phi+X\phi^2\right)\,,
\label{bosop}
\ee
where the {\it{mass}} term $X$ can be an arbitrary $\sigma$-dependent function.

If we define the measure in the scalar functional space to be scaled with an arbitrary function $\mathcal{M}$
\be
\left(\phi,\phi^{\prime}\right)=\int d^2\sigma\sqrt{g}\mathcal{M}\phi\phi^{\prime}\,,
\label{bosmeasure}
\ee
the relevant determinant when computing the path integral results
\be
\det{\Delta^B_{\mathcal{M}}}=\det{\mathcal{M}^{-1}(\nabla^2_g+X)}\,,
\ee
where $\nabla^2_g$ is the scalar Laplacian in the arbitrary metric $g_{ij}$.

The conformal anomaly under this rescaling arises in the heat kernel expansion for these operator
\be
\log\det{\Delta^B_{\mathcal{M}}}=-\int^{\infty}_{\Lambda^{-2}} \frac{dt}{t}\textrm{Tr}\exp{\left(-t\Delta^B_{\mathcal{M}}\right)}\,,
\ee
where $\Lambda^{-2}$ is an appropriate cut-off since the integral is divergent  $t\to0$ . Making use of $\delta\Delta^B_{\mathcal{M}}=-\delta\log{\mathcal{M}}\Delta^B_{\mathcal{M}}$ we obtain
\be
\delta\log\det{\Delta^B_{\mathcal{M}}}=\textrm{Tr}\left(\exp{\left(-t\Delta^B_{\mathcal{M}}\right)}\delta\log\mathcal{M}\right)\mid^{\infty}_{\Lambda^{-2}}\,.
\label{vardet}
\ee
The $t\to\infty$ limit vanishes for positive definite operators, so the only contribution comes from the $t\to 0$ limit. In this limit we expand the heat kernel as follows
\be
\textrm{Tr}\left(\exp{\left(-t\Delta^B_{\mathcal{M}}\right)}\delta\log\mathcal{M}\right)\sim\frac{1}{t^{d/2}}\sum_n a_n\left(\delta{\log{\mathcal{M}}}\mid\Delta^B_{\mathcal{M}}\right) t^n\,,
\ee
where $d$ is dimension of the manifold and $a_n$ are the Seeley coefficients. Since the manifold in our problem has a boundary, $n$ corresponds to integer and semi-integer positive numbers.

For $d=2$, we have divergent contributions proportional to $\Lambda^2$, $\Lambda$ and $\log\Lambda$ which involve $a_0$, $a_{1/2}$ and $a_1$. It can be seen  that these terms do not depend on $\mathcal{M}$, \cite{Schwarz:1979ae,DGT}. The relevant contribution comes from the finite term which depends only on $a_1$ (and irrelevant boundary terms). Then we conclude that
\be
\delta\log\det{\Delta^B_{\mathcal{M}}}=-a_1\left(\delta{\log{\mathcal{M}}}\mid\Delta^B_{\mathcal{M}}\right)\,.
\label{variation1}
\ee
The Seeley coefficient  $a_1$ for these operators is known. Therefore, we obtain the conformal anomaly in terms of a Liouville type action for $\mathcal{M}$
\be
\log\det{\Delta^B_{\mathcal{M}}}-\log\det{\Delta^B} =
-\frac{1}{4\pi}\int d^2\sigma\sqrt{g}\left(\log\mathcal{M}\left(\frac16R^{(2)}-X\right)+\frac{1}{12}\partial^i\log\mathcal{M}\partial_i\log\mathcal{M}\right)\,,
\label{scalaranomaly}
\ee
where $R^{(2)}$ is the scalar curvature for the metric $g_{ij}$.
When $\mathcal{M}=\frac{1}{\sqrt{g}}$ we can integrate by parts and rewrite (\ref{scalaranomaly}) in the form
\be
\log\det{\Delta^B_{\mathcal{M}}}-\log\det{\Delta^B} =
+\frac{1}{4\pi}\int d^2\sigma\sqrt{g}\left(\log\mathcal{M}X+\frac{1}{12}\partial^i\log\mathcal{M}\partial_i\log\mathcal{M}\right)\,.
\label{scalaranomaly2}
\ee

~

Let us now consider the action for a fermionic operator $D_F$
\be
\int d^2\sigma\sqrt{g}\bar{\psi}D_F\psi\,,
\ee
and measure
\be
\left(\psi,\psi^{\prime}\right)=\int d^2\sigma\sqrt{g}\mathcal{K}\bar{\psi}\psi^{\prime}\,.
\label{fermmeasure}
\ee
In this case the relevant second order operator is $\Delta^F_{\mathcal{K}}=(\mathcal{K}^{-1}D_F)^2$, whose variation is
\be
\delta\log\Delta^F_{\mathcal{K}}= \int^{\infty}_{\Lambda^{-2}} dt \textrm{Tr}\exp{\left(-t\Delta^F_{\mathcal{K}}\right)}\left(\delta\mathcal{K}^{-1}D_F\mathcal{K}^{-1}D_F+\mathcal{K}^{-1}D_F\delta\mathcal{K}^{-1}D_F\right)\,.
\ee
Because of the trace, both terms commute and we obtain
\be
\delta\log\Delta^F_{\mathcal{K}}= -2 \textrm{Tr}\exp{\left(-t\Delta^F_{\mathcal{K}}\right)}\delta\log\mathcal{K}\mid^{\infty}_{\Lambda^{-2}}\,.
\ee
As done in the bosonic case, we can compute the anomaly in terms of the Seeley coefficients of $\Delta^F_{\mathcal{K}}$
\be
\delta\log\det{\Delta^F_{\mathcal{K}}}=-2a_1\left(\delta{\log{\mathcal{K}}}\mid\Delta^F_{\mathcal{K}}\right)\,.
\label{variation2}
\ee
Note that for non-trivial $\mathcal{K}$ this is in general different from $\mathcal{K}^{-2}D_F^2$ because of the linear derivatives acting on $\mathcal{K}^{-1}$. Therefore we have to compute the corresponding Seeley coefficients for $(\mathcal{K}^{-1}D_F)^2$ which are in general different from the $\mathcal{K}^{-2}D_F^2$ ones.

Consider
\be
D_F^2=-\hat{\nabla}^2+\frac14R^{(2)}+Y\,,
\ee
where $\hat{\nabla}$ involves spin connection terms and $Y$ is an arbitrary potential. For this class of operators the Seeley coefficient $a_1$ is also known \cite{DGT}. Then, the anomaly results
\be
\log\det{\Delta^F_{\mathcal{K}}}-\log\det{\Delta^F} =
\frac{1}{4\pi}\int d^2\sigma\sqrt{g}\left(\log\mathcal{K}\left(\frac16R^{(2)}+2Y\right)+\frac{1}{6}\partial^i\log\mathcal{K}\partial_i\log\mathcal{K}\right)\,,
\label{fermranomaly}
\ee
and for $\mathcal{K}=\frac{1}{(\sqrt{g})^{1/2}}$ we can rewrite it as
\be
\log\det{\Delta^F_{\mathcal{K}}}-\log\det{\Delta^F} =
\frac{1}{4\pi}\int d^2\sigma\sqrt{g}\left(\log\mathcal{K}(2Y)-\frac{1}{6}\partial^i\log\mathcal{K}\partial_i\log\mathcal{K}\right)\,.
\label{fermranomaly2}
\ee
The last expression is valid for general 2d fermionic operators. However, the in the GS formulation the fermionic fields are worldsheet anticommuting scalars. In this case, the $R^{(2)}$ term is four times the corresponding to standard 2d fermions \cite{DGT}. Taking this into account we obtain
\be
\log\det{\Delta^F_{\mathcal{K}}}-\log\det{\Delta^F} =
\frac{1}{4\pi}\int d^2\sigma\sqrt{g}\left(\log\mathcal{K}(2Y)-\frac{2}{3}\partial^i\log\mathcal{K}\partial_i\log\mathcal{K}\right)\,.
\label{GSfermranomaly2}
\ee

Now that we know the individual contribution of scalars and GS fermions to the scaling anomaly, we are able to derive the condition over the partition function in order to be invariant under this operation.

For that purpose, consider the fluctuation scalar fields over a classical configuration. In general, the action can be written in the form
\be
S^B_{1-loop}=\int d^2\sigma\sqrt{g}\left(g^{ij}\partial_i\phi^a\partial_j\phi^a + X_{ab}\phi^a\phi^b\right)\,,
\ee
where summation over $a=1,\ldots,10$ is implied and $g_{ij}$ is the induced metric  for a general classical configuration. All the possible potential and mass terms are contained in the {\it mass} matrix $X_{ab}$. Since we are now working in  the conformal gauge, this fluctuation Lagrangian contains both transverse and longitudinal scalar modes.

The fluctuation Lagrangian for the fermionic modes can be written as
\be
S^F_{1-loop}\int d^2\sigma\sqrt{g}\bar{\psi}^a(D_F)_{ab}\psi^b\,,
\ee
where $a=1,\ldots,8$ runs over the 8 two component GS fermionic modes. Then the quadratic fermionic operators are
\be
(D_F)_{(ab)}^2\psi^b=\left(D^2\delta_{ab}+Y_{ab}\right)\psi^b\,,
\ee
where $D^2$ is the square of the standard kinetic term for each GS fermion and the {\it mass} matrix $Y_{ab}$ contains all possible potential and mass terms.

Finally, for the ghost modes with $\mathcal{K}=\left(\sqrt{g}\right)^{-1/2}$, the anomaly can be written as
\be
\log\det{\Delta^{gh}_{\mathcal{K}}}-\log\det{\Delta^{gh}} =
-\frac{26}{12}\frac{1}{4\pi}\int d^2\sigma\sqrt{g}\partial^i\log\mathcal{K}\partial_i\log\mathcal{K}\,.
\label{ghostanomaly}
\ee
Given the scaling anomalies (\ref{scalaranomaly2}), (\ref{GSfermranomaly2}) and (\ref{ghostanomaly})
we can see \cite{DGT} that the cancellation of the total scaling anomaly is equivalent to the condition
\be
{\textrm{Tr}} X = {\textrm{Tr}} Y\,.
\label{cond}
\ee
The results presented in Appendix \ref{classical} were computed in \cite{forini} in static gauge (Nambu-Goto formulation). In order to verify the last equality for this configuration note that, being $\bar{X}$ the {\it mass} term obtained in static gauge, it is related to the corresponding one in conformal gauge as
\be
\textrm{Tr}\bar{X}=\textrm{Tr}X-\textrm{Tr}^{long}X\,.
\label{ngtr}
\ee
Since it is known that $\textrm{Tr}^{long}X=-R^{(2)}$, the condition \eqref{cond} can be expressed as
\be
{\textrm{Tr}} \bar{X} - {\textrm{Tr}} Y = R^{(2)}\,.
\label{cond2}
\ee
It is important to stress that this scaling invariance condition is valid for any string configuration. In particular, for the straight line configuration ($R^{(2)}=-2$) we have a 2d field theory on $AdS_2$ with 2 fermionic massless modes, 6 with $Y_{aa}=m_F^2=1$ and
\be
\bar{X}_{ab}= {\textrm{diag}}\left(R^{(2)}+4,2,0,0,0,0,0,0\right)\,.
\ee
Therefore condition \eqref{cond2} is satisfied and we can compute the 1-loop partition function via a vacuum energy computation.

Finally, for the cusped classical configuration the condition \eqref{cond2} is also valid implying the vanishing of the total scaling anomaly and therefore justifying the vacuum energy computation made in this article. Note that this analysis is valid for arbitrary values of parameters $p$ and $q$.

%%%%%%%%%%%%%%%%%%%%%%%%%%%%%%%%%%%%%%%%%%%%%%%%%%%%%%%%%%%%%%%%%%%%%%%%%%%%%%%%%%%%%%%%%%%%%%%%%%%%%%%%


\begin{thebibliography}{99}
%%%%%%%%%%%%%%%%%%%%%%%%%%%%%%%%%%%%%%%%%%%%%%%%%%%%%%%%%%%%%%%%%%%%%%%%%%%%%%%%%%%%%%%%%%%%%%%%%%%%%%%%


\bibitem{Correa:2012at}
  D.~Correa, J.~Henn, J.~Maldacena and A.~Sever,
  %``An exact formula for the radiation of a moving quark in N=4 super Yang Mills,''
  JHEP {\bf 1206} (2012) 048
  [arXiv:1202.4455 [hep-th]].

\bibitem{Fiol:2012sg}
  B.~Fiol, B.~Garolera and A.~Lewkowycz,
  %``Exact results for static and radiative fields of a quark in N=4 super Yang-Mills,''
  JHEP {\bf 1205} (2012) 093
  [arXiv:1202.5292 [hep-th]].



\bibitem{Lewkowycz:2013laa}
 A.~Lewkowycz and J.~Maldacena,
  %``Exact results for the entanglement entropy and the energy radiated by a quark,''
  JHEP {\bf 1405} (2014) 025
  [arXiv:1312.5682 [hep-th]].

\bibitem{Drukker:2011za}
  N.~Drukker and V.~Forini,
  %``Generalized quark-antiquark potential at weak and strong coupling,''
  JHEP {\bf 1106} (2011) 131
  [arXiv:1105.5144 [hep-th]].


\bibitem{abjm}
 O.~Aharony, O.~Bergman, D.~L.~Jafferis and J.~Maldacena,
 %``N=6 superconformal Chern-Simons-matter theories, M2-branes and their gravity duals,''
 JHEP {\bf 0810} (2008) 091
 [arXiv:0806.1218 [hep-th]].

\bibitem{Griguolo:2012iq}
  L.~Griguolo, D.~Marmiroli, G.~Martelloni and D.~Seminara,
  %``The generalized cusp in ABJ(M) N = 6 Super Chern-Simons theories,''
  JHEP {\bf 1305} (2013) 113  [arXiv:1208.5766 [hep-th]].

\bibitem{Cardinali:2012ru}
  V.~Cardinali, L.~Griguolo, G.~Martelloni and D.~Seminara,
  %``New supersymmetric Wilson loops in ABJ(M) theories,''
  Phys.\ Lett.\ B {\bf 718} (2012) 615
  [arXiv:1209.4032 [hep-th]].

\bibitem{forini}
  V.~Forini, V.~Giangrecco~M.~Puletti, O.~Ohlsson~Sax,
  %``The generalized cusp in AdS4 x CP3 and more one-loop results from semiclassical strings,''
  J. Phys.  {\bf A 46 (2013) 115402}
  arXiv:1204.3302 [hep-th].

\bibitem{forini2loop}
    L.~Bianchi, M.~Bianchi, A.~Brès, V.~Forini, E.~Vescovi,
    %``Two-loop cusp anomaly in ABJM at strong coupling,''
    JHEP {\bf 1410} (2014) 013
    arXiv:1407.4788 [hep-th].


\bibitem{Drukker:2009hy}
  N.~Drukker and D.~Trancanelli,
  %``A Supermatrix model for N=6 super Chern-Simons-matter theory,''
  JHEP {\bf 1002} (2010) 058
  [arXiv:0912.3006 [hep-th]].


\bibitem{Bianchi:2014laa}
  M.~S.~Bianchi, L.~Griguolo, M.~Leoni, S.~Penati and D.~Seminara,
  %``BPS Wilson loops and Bremsstrahlung function in ABJ(M): a two loop analysis,''
  JHEP {\bf 1406} (2014) 123  [arXiv:1402.4128 [hep-th]].


\bibitem{Correa:2014aga}
  J.~Aguilera-Damia, D.~H.~Correa and G.~A.~Silva,
  %``Strings in $AdS_4 \times \mathbb{CP}^{3}$ Wilson loops in $\mathcal N=$6 super Chern-Simons-matter and bremsstrahlung functions,''
  JHEP {\bf 1406} (2014) 139 arXiv:1405.1396 [hep-th].


\bibitem{Kapustin:2009kz}
  A.~Kapustin, B.~Willett and I.~Yaakov,
  %``Exact Results for Wilson Loops in Superconformal Chern-Simons Theories with Matter,''
  JHEP {\bf 1003} (2010) 089
  [arXiv:0909.4559 [hep-th]].

\bibitem{Drukker:2010nc}
  N.~Drukker, M.~Marino and P.~Putrov,
  %``From weak to strong coupling in ABJM theory,''
  Commun.\ Math.\ Phys.\  {\bf 306} (2011) 511
  [arXiv:1007.3837 [hep-th]].


\bibitem{Klemm:2012ii}
  A.~Klemm, M.~Marino, M.~Schiereck and M.~Soroush,
  %``ABJM Wilson loops in the Fermi gas approach,''
  arXiv:1207.0611 [hep-th].


\bibitem{Camporesi:1992wn}
  R.~Camporesi and A.~Higuchi,
  %``Stress energy tensors in anti-de Sitter space-time,''
  Phys.\ Rev.\ D {\bf 45} (1992) 3591.


\bibitem{Schwarz:1979ae}
  A.~S.~Schwarz,
  %``The Partition Function of a Degenerate Functional,''
  Commun.\ Math.\ Phys.\  {\bf 67} (1979) 1.

\bibitem{DGT}
  N.~Drukker, D.~J.~Gross and A.~A.~Tseytlin,
  %``Green-Schwarz string in AdS(5) x S**5: Semiclassical partition function,''
  JHEP {\bf 0004} (2000) 021
  [hep-th/0001204].


\bibitem{Sakai:1984vm}
  N.~Sakai and Y.~Tanii,
  %``Supersymmetry in Two-dimensional Anti-de Sitter Space,''
  Nucl.\ Phys.\ B {\bf 258} (1985) 661.


\bibitem{DPY}
  N.~Drukker, J.~Plefka and D.~Young,
  %``Wilson loops in 3-dimensional N=6 supersymmetric Chern-Simons Theory and their string theory duals,''
  JHEP {\bf 0811} (2008) 019
  [arXiv:0809.2787 [hep-th]].

\bibitem{Chen:2008bp}
  B.~Chen and J.~-B.~Wu,
  %``Supersymmetric Wilson Loops in N=6 Super Chern-Simons-matter theory,''
  Nucl.\ Phys.\ B {\bf 825} (2010) 38
  [arXiv:0809.2863 [hep-th]].

\bibitem{Rey:2008bh}
  S.~-J.~Rey, T.~Suyama and S.~Yamaguchi,
  %``Wilson Loops in Superconformal Chern-Simons Theory and Fundamental Strings in Anti-de Sitter Supergravity Dual,''
  JHEP {\bf 0903} (2009) 127
  [arXiv:0809.3786 [hep-th]].


\bibitem{Forini:2010ek}
  V.~Forini,
  %``Quark-antiquark potential in AdS at one loop,''
  JHEP {\bf 1011} (2010) 079
  [arXiv:1009.3939 [hep-th]].

\bibitem{Beccaria:2010ry}
  M.~Beccaria, G.~V.~Dunne, V.~Forini, M.~Pawellek and A.~A.~Tseytlin,
  %``Exact computation of one-loop correction to energy of spinning folded string in AdS_5 x S^5,''
  J.\ Phys.\ A {\bf 43} (2010) 165402
  [arXiv:1001.4018 [hep-th]].

\bibitem{Alvarez:1982zi}
  O.~Alvarez,
  %``Theory of Strings with Boundaries: Fluctuations, Topology, and Quantum Geometry,''
  Nucl.\ Phys.\ B {\bf 216} (1983) 125.


\bibitem{Luckock:1989mv}
  H.~Luckock,
  %``Quantum Geometry of Strings With Boundaries,''
  Annals Phys.\  {\bf 194} (1989) 113.


\end{thebibliography}
\end{document}